\documentclass{aa}

\usepackage[dvips]{graphics}
\usepackage{txfonts}

\def\WAT{H$_2$O}

\def\hdo {H$_2$O}
\def\Lfir{$L_{\rm fir}$}
\def\Lmax{$L_{\rm H_2O}^{\rm max}$}
\def\Lup{$L_{\rm H_2O}^{\rm up}$}
\def\vlsr{$V_{\rm lsr}$}

\newcommand{\lsol}{L{$_{\odot}$}}

\newcommand{\kms}{km~s{$^{-1}$}}
\newcommand{\kmsyr}{km~s{$^{-1}$\,yr$^{-1}$}}

\newcommand{\gsim}{\;\lower.6ex\hbox{$\sim$}\kern-7.75pt\raise.65ex\hbox{$>$}\;}
\newcommand{\lsim}{\;\lower.6ex\hbox{$\sim$}\kern-7.75pt\raise.65ex\hbox{$<$}\;}

\begin{document}
\title{Long-term study of water masers associated with Young
Stellar Objects. II: Analysis}

\author{J. Brand\inst{1}, R. Cesaroni\inst{2}, G. Comoretto\inst{2}, 
M. Felli\inst{2}, F. Palagi\inst{3}, F. Palla\inst{2}, 
and R. Valdettaro\inst{2}}

\offprints{J. Brand, \email{brand@ira.cnr.it}}

\institute{
Istituto di Radioastronomia C.N.R., Via Gobetti 101, I-40129 Bologna, Italy
\and
INAF - Osservatorio Astrofisico di Arcetri, Largo E. Fermi 5, I-50125 Firenze, 
Italy
\and
Istituto di Radioastronomia C.N.R., Sezione Firenze, 
Largo E. Fermi 5, I-50125 Firenze, Italy}

\date{Received; Accepted }

\abstract{We present the analysis of the properties of water maser emission
in 14 star forming regions (SFRs), which have been monitored for up to 13
years with a sampling rate of about once every 2-3 months. The 14 regions were 
chosen to span a range in luminosity \Lfir\ of the associated Young
Stellar Object (YSO) between 20~\lsol\ and $1.8 \times 10^6$~\lsol.
The general scope of the analysis is to investigate the dependence of the 
overall spectral morphology of the maser emission and its variability on the
luminosity of the YSO. We find that higher-luminosity sources tend to
be associated with stronger and more stable masers. Higher-luminosity YSOs
can excite more emission components over a larger range in velocity, yet
the emission that dominates the spectra is at a velocity very near that of the
molecular cloud in which the objects are embedded.
For \Lfir\ $\gsim 3 \times 10^4$~\lsol\ the maser emission
becomes increasingly structured and more extended in velocity with increasing
\Lfir. Below this limit the maser emission shows the same variety of 
morphologies, but without a clear dependence on \Lfir\ and with a smaller
velocity extent. Also, for sources with \Lfir\ above
this limit, the water maser is always present above the 5$\sigma$-level; below 
it, the typical 5$\sigma$ detection rate is 75-80\%. 
Although the present sample contains few objects
with low YSO luminosity, we can conclude that there must be a lower limit to
\Lfir\ ($\lsim 430$~\lsol), below which the associated maser is below the 
detection level most of the time. 
These results can be understood in terms of scaled versions of similar SFRs 
with different YSO luminosities, each with 
many potential sites of maser amplification, which can be excited provided
there is sufficient energy to pump them, i.e. the basic pumping process is 
identical regardless of the YSO luminosity. In SFRs with lower input 
energies, the conditions of maser amplification are much closer to the 
threshold conditions, and consequently more unstable.

\noindent
We find indications that the properties of the maser emission may be determined
also by the geometry of the SFR, specifically by the beaming and collimation
properties of the outflow driven by the YSO.

\noindent
For individual emission components the presence of
velocity gradients seems to be quite common; we find both
acceleration and deceleration, with values between 0.02 and 1.8~\kmsyr.

\noindent
From the 14 `bursts' that we looked at in some detail we derive 
durations of between 60 and 900~days and flux density increases of between 
40\% and $\gsim 1840$\% (with an absolute maximum of $\sim 820$~Jy over 63 
days). 
The ranges found in burst- intensity and -duration are biased by our minimum 
sampling interval, while the lifetime of the burst is furthermore affected by 
the fact that bursts of very long duration may not be recognized as such. 

\noindent
In addition to the flux density variations in individual emission components, 
the \WAT\ maser output 
as a whole is found to exhibit a periodic long-term variation in several 
sources. This may be a consequence of periodic variations in the wind/jets 
from the exciting YSO.
}

\titlerunning{Water maser variability}
\authorrunning{J. Brand et al.}
\maketitle

\section{Introduction}

In Paper~I of this series (Valdettaro  et al.~\cite{paper1}) we presented the 
results of more than 10 years of single-dish monitoring of the \WAT\ maser
emission in 14 Star Forming Regions (SFRs) obtained with the Medicina 32-m
radiotelescope. The average time interval between two successive observations
is 2$-$3 months, so that our database allows one
to study only the long-term ($>$ 2$-$3 months) aspects of the maser
variability. For each source a brief description of the maser
environment was given in Paper~I, emphasizing what is of 
interest in the discussion of the \WAT\ maser variability.
Here we present quantitative results derived from the long-term
monitoring. 

The amount of information collected on the \WAT\ maser emission during
our study is exceedingly large. To compress it to a more manageable 
form the following quantities were derived (see also Paper~I):

\begin{enumerate}
\item The flux density $F$ as a function of both velocity and time, 
presented in velocity-time-intensity plots, which give the best overall 
description of the maser activity and 
help to visually identify possible velocity drifs of the emission;

\item The integrated flux density $S$ as a function of time, which describes 
the variation of the total maser emission;
 
\item The upper and the lower envelopes of the spectra over the whole period
of observation, obtained by assigning to each velocity channel respectively the
maximum (if $> 5\sigma$), and minimum (=0, unless it's $> 5\sigma$) signal 
detected during the monitoring period;

\item The potential maximum maser luminosity \Lup, derived by
integration of the upper envelope. This quantity represents the maximum output 
which the source could produce {\it if} all the velocity components were to
emit {\it at their maximum level} and {\it at the same time}; 

\item The actually {\it observed} maximum maser luminosity \Lmax, derived 
from the spectrum with the highest integrated flux density;

\item The frequency of occurrence of a spectral feature. To produce these plots
the spectra were re-binned with a velocity resolution
$\Delta V \simeq 0.3$ km s$^{-1}$, and for each channel a counter 
was increased by one every time the flux density in that channel
was greater than 5$\sigma$;

\item The first moment of the upper envelope, $V_{\rm up}$, i.e. the average
velocity weighted by the flux density, with its second moment 
$\Delta V_{\rm up}$, and the first moment of the frequency-of-occurrence, 
$V_{\rm fr}$, i.e. the average velocity weighted by the number of times that 
velocity component is present in the spectra, with its second moment 
$\Delta V_{\rm fr}$.

\end{enumerate}

\noindent
Since one of our main aims is to reveal aspects of 
the \WAT\ maser variability that depend on the luminosity of the
Young Stellar Object (YSO) exciting the maser, the selected sample 
covers rather uniformly a large range of (FIR) luminosities, 
from 20 L$_\odot$ to 1.8 10$^6$ L$_\odot$, 
which brackets almost the entire luminosity interval of the exciting sources
of \WAT\ masers in SFRs
\footnote{VLBI observations (e.g. Seth et al.~\cite{seth}) show that the
presence of several distinct maser groups (i.e. YSOs) in a SFR is more the
rule than the exception. \Lfir\ is a global parameter that accounts for the
emission of all YSOs and cannot be directly related to any of the maser
groups that might be present. Only VLBI maser observations combined with
very high-resolution FIR studies will be able to associate to each maser group
its own \Lfir. In the present context \Lfir\ is used to set an upper limit
to the luminosity of the brightest YSO in the SFR and in this sense allows
discrimination between the environments of YSOs of different luminosities.}
(Palagi et al.~\cite{PCC93}; Wilking et al.~\cite{wilking}; 
Furuya et al.~\cite{furuya01}, \cite{furuya03}). 

\noindent
In Table~\ref{tsample} we present the source sample as described in Paper~I,
ordered in terms of increasing FIR luminosity: 
Col.~1 gives the sequential number as given in
Table~1 of Paper~I; Cols.~ 2 and 3 give source names and the associated
IRAS source; Cols.~4 and 5 list the B1950 coordinates; in Cols.~6
to 8 we give the radial velocity relative to the LSR ($V_{\rm cl}$) of the 
molecular cloud in which the SFR is embedded, the distance ($d$), and the FIR 
luminosity (\Lfir) of the source; detailed references for these
quantities are given in Table~1 of Paper~I. Cols.~9 to 12 list: $V_{\rm fr}$,
$\Delta V_{\rm fr}$, $V_{\rm up}$, and $\Delta V_{\rm up}$ respectively, while 
in Col.~13 we give the total integrated H$_2$O flux density,
determined from the upper envelope, with the corresponding potential maximum 
maser luminosity
\Lup\ in Col.~15. Finally, Cols.~14 and 16 give the maximum
integrated H$_2$O flux density measured during the monitoring campaign, and 
the corresponding H$_2$O luminosity \Lmax, respectively. 

\noindent
{\bf Note:} After publication of Paper~I, we found that the gain curve used
for the May 12, 1998 data was in error. We have therefore recalibrated the
spectra taken on that date. Depending on source elevation, the new flux 
densities may be up to a factor of 3-4 smaller than what was reported 
previously. In
Table~\ref{newmay98} we give the new parameters of the affected spectra.

\noindent
In the present paper we shall analyse the properties of the maser emission
in general terms, without discussing individual sources in any great 
detail. For particularly interesting sources, more in-depth studies may be 
presented in separate papers in the future.

\begin{table*}
\caption{ The \hdo\ maser sample, arranged in order of increasing FIR
luminosity. Main derived parameters.}
\label{tsample}
\resizebox{18cm}{!}{\rotatebox{90}{\includegraphics{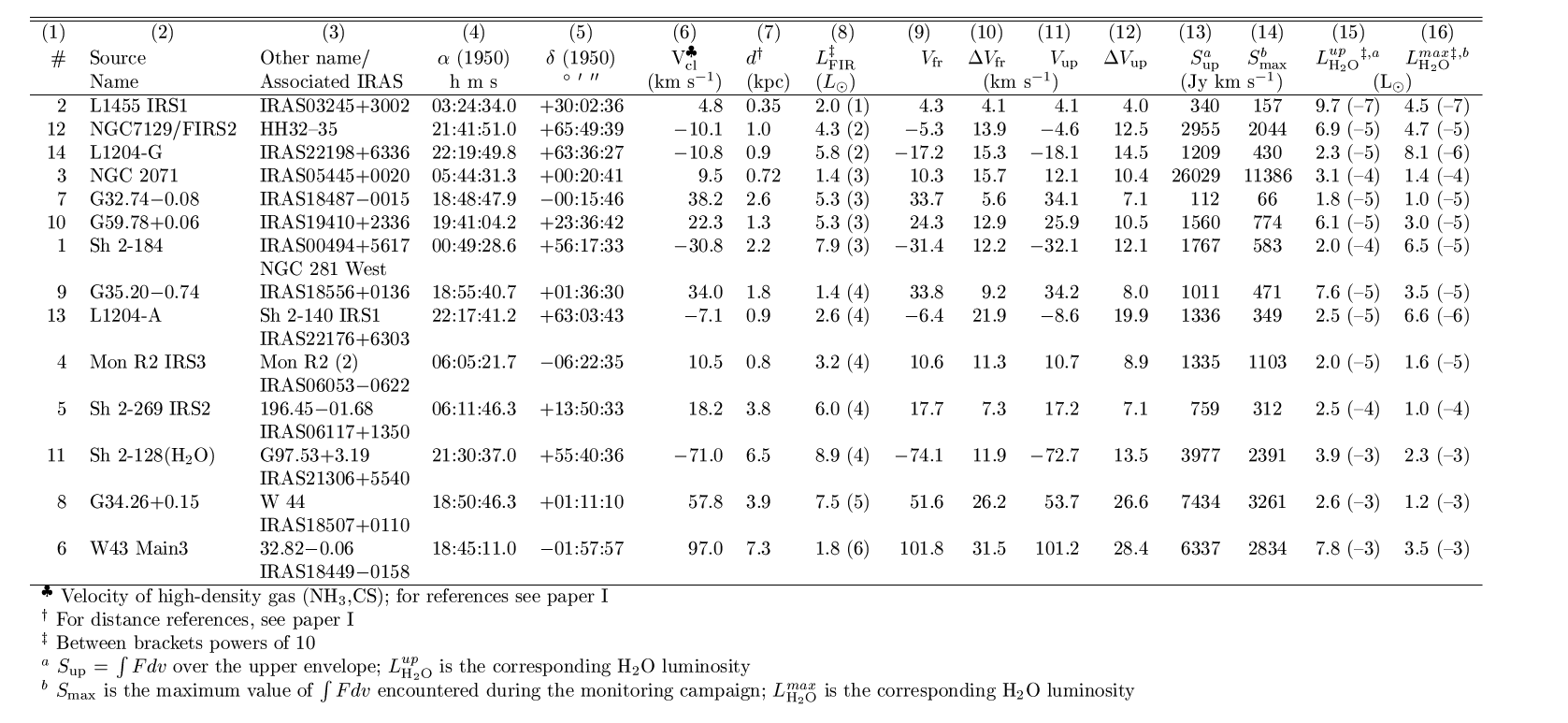}}}
\end{table*}

\begin{table}
\caption[]{\label{newmay98} Recalibrated data for the May 12, 1998 spectra}
\resizebox{9cm}{!}{\includegraphics{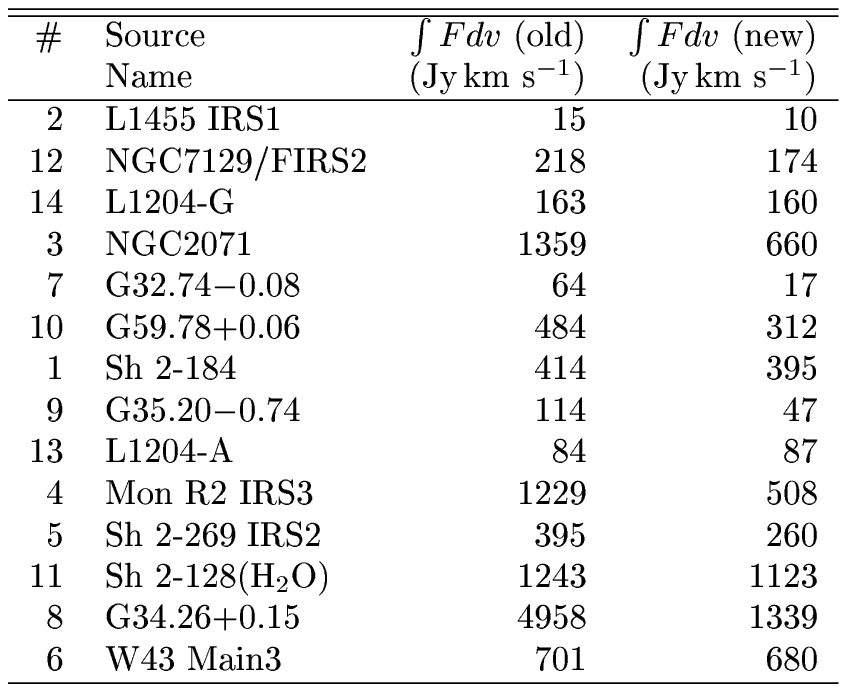}}
\end{table}

\section{General Results}

\begin{figure*}
\resizebox{12cm}{!}{\rotatebox{270}{\includegraphics{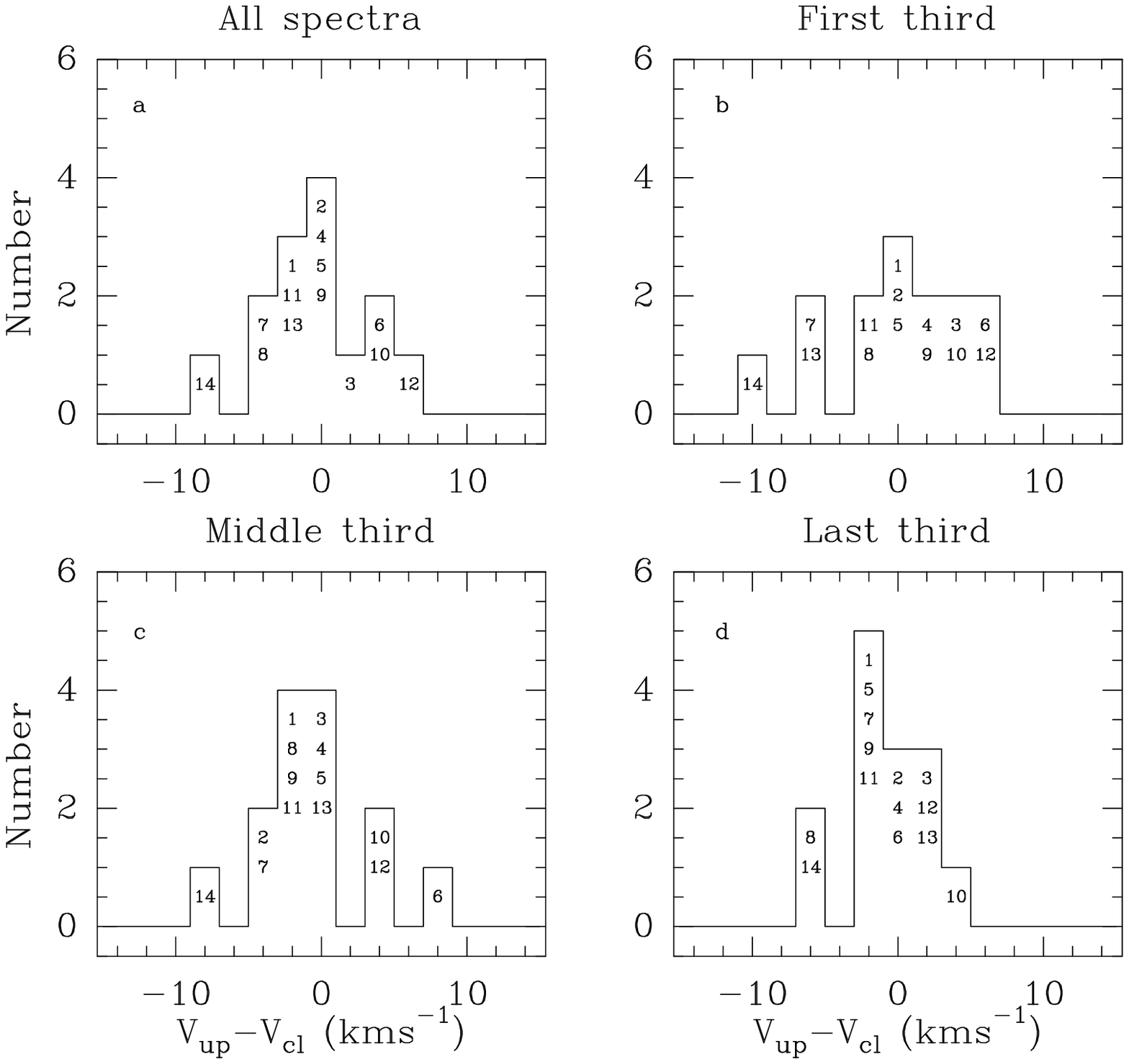}}}
\caption[]{Distribution of the velocity difference between the mean velocity
($V_{\rm up}$) of the ``upper envelope'' spectrum and the molecular cloud
velocity ($V_{\rm cl}$), where the ``upper envelope'' was created from:
{\bf a}\ all spectra, {\bf b} the first third, {\bf c}\ the middle third, and
{\bf d}\ the last third of all spectra, respectively. The identification 
numbers of the sources (see Tab~\ref{tsample}) in each bin are indicated.}
\label{vhisto}
\end{figure*}

\subsection{Mean velocities}

In the literature the velocity that defines the maser emission is usually
taken to be that of the peak in the spectrum. More rarely, in the presence of 
very complex spectra with multiple peaks, an average or centroid velocity is 
used. Given the large changes in the spectra when observed over long periods
of time, both of these definitions tend to be a function of the observing date.
Consequently it is not a surprise that the velocity of the maser, as quoted in 
the literature, may change with time. 

\noindent
These maser velocities are usually compared with those of the molecular 
clouds in which the SFRs are embedded. The results indicate a good
general agreement between the two, with a dispersion $\simeq 4-11$~\kms\ 
(Wouterloot et al.~\cite{wou95}, Anglada et al.~\cite{ang96}).

\noindent
Our first-moment velocities $V_{\rm up}$ and $V_{\rm fr}$ are derived from
the upper envelope and the frequency-of-occurrence plots, respectively, and 
refer to a long
($\sim 10$~yrs) period of observation. Consequently they offer a way to define 
a mean velocity of the maser emission that is less dependent on the epoch of 
the observation.

\noindent
As indicated in Table~\ref{tsample}, $V_{\rm up}$ and $V_{\rm fr}$ are almost 
identical, with a mean difference of 0.2~km\,s$^{-1}$ and a standard
deviation of 1.2~km\,s$^{-1}$. {\it This implies that the velocity at which 
the emission is most intense is also that where emission occurs most 
frequently}.

\begin{figure*}[tp]
\resizebox{7.5cm}{!}{\rotatebox{270}{
\includegraphics{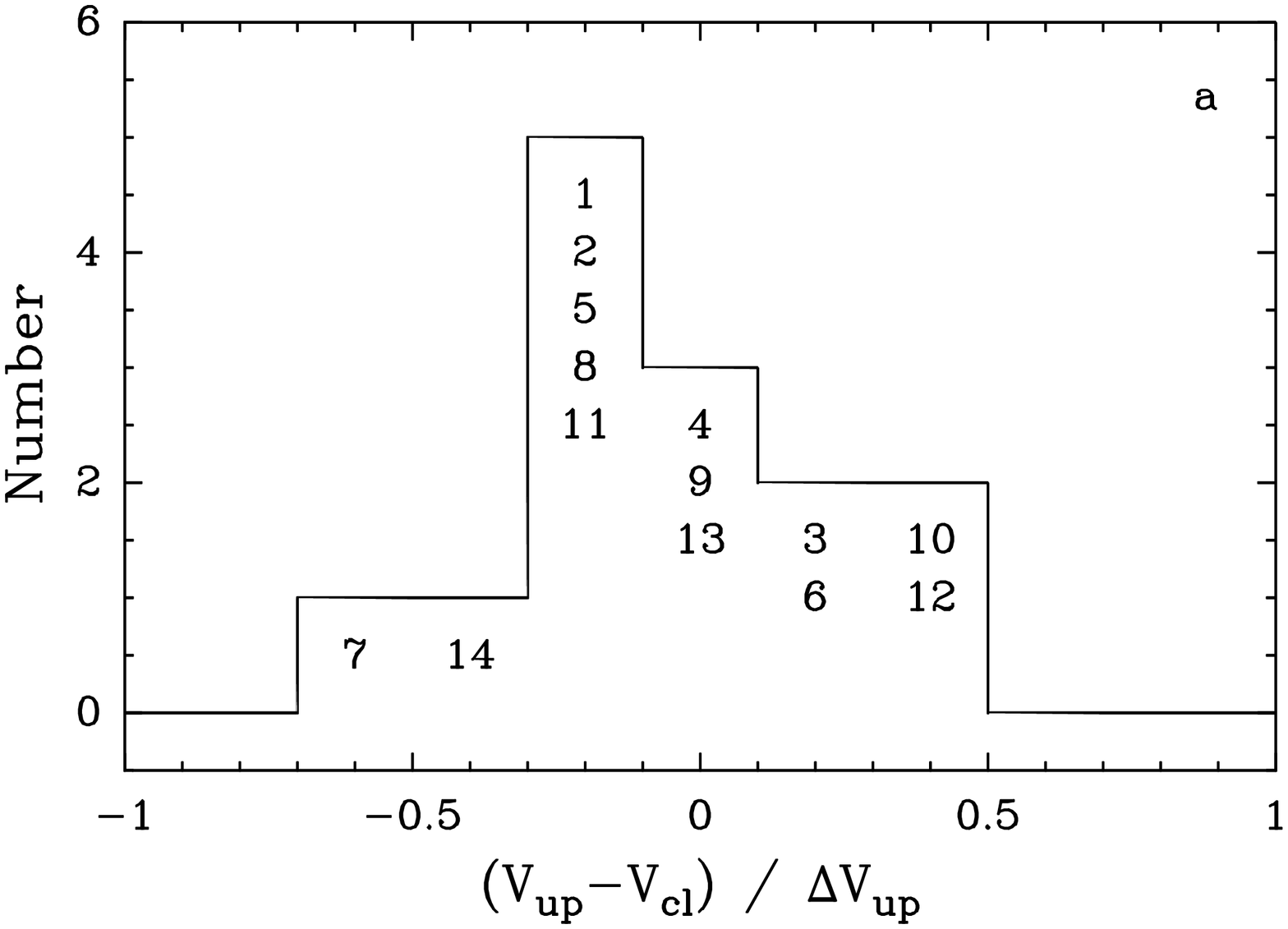}}}
\hspace{1cm}
\noindent
\resizebox{7.5cm}{!}{\rotatebox{270}{
\includegraphics{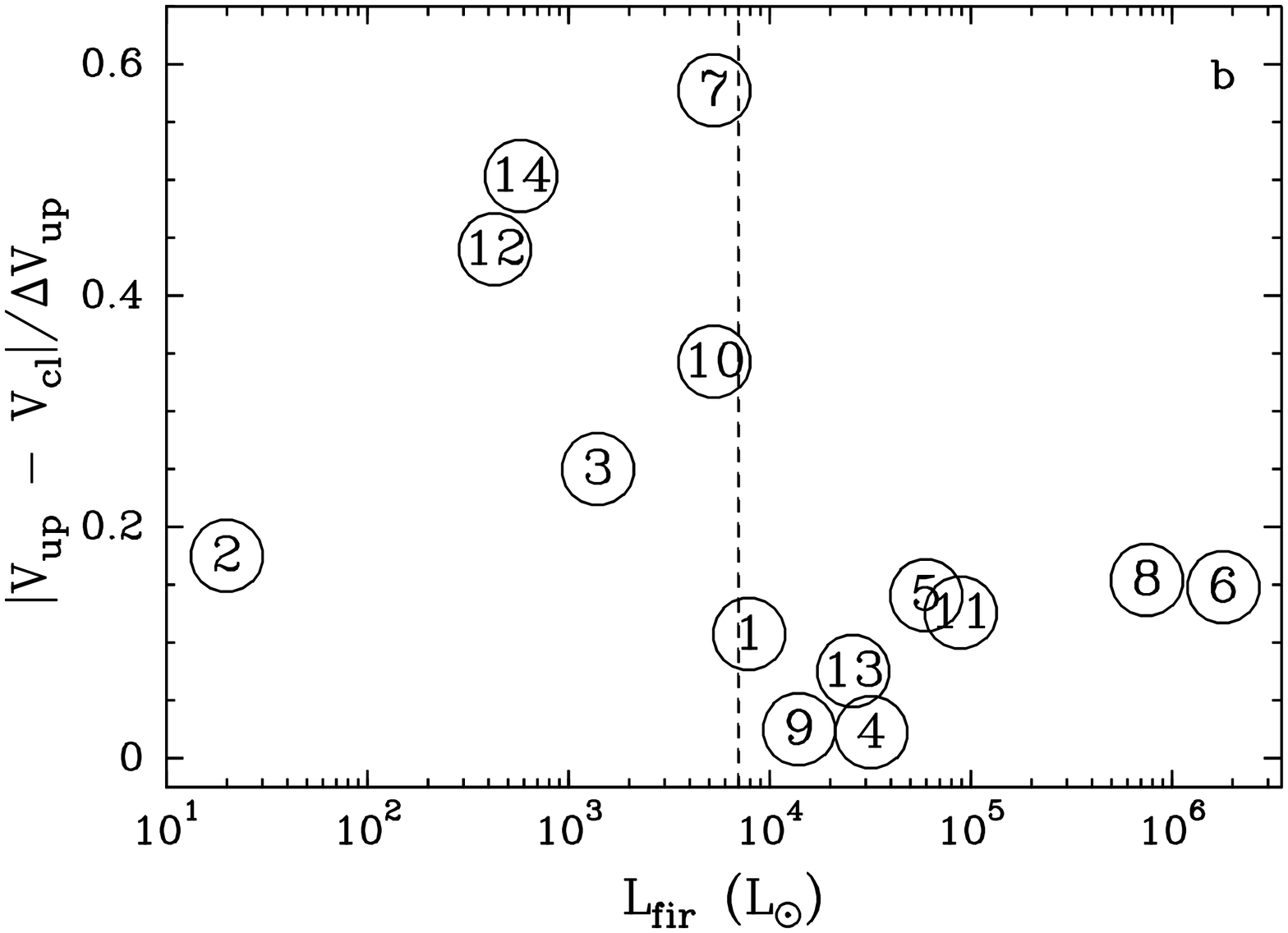}}}
\caption[]{{\bf a}\ Distribution of the velocity difference between the mean
velocity of the ``upper envelope'' spectrum and the molecular cloud velocity,
relative to the velocity dispersion (second moment) of the maser emission
($\Delta V_{\rm up}$). The identification numbers of the sources in each bin
are indicated. {\bf b}\ The absolute value of the parameter shown in {\bf a},
as a function of the FIR luminosity of the associated YSO. The dashed line 
marks \Lfir = $3.7 \times 10^3$~\lsol, above which the relative velocity 
difference between maser and cloud seems to be smaller (see text).}
\label{vhistofrac}
\end{figure*}

\noindent
It is also instructive to compare $V_{\rm up}$ or $V_{\rm fr}$ with the
velocity  of the molecular cloud. As shown in Fig.~\ref{vhisto}a,
$|V_{\rm up}|$ differs always less than $\sim 7.5$~\kms\ from the 
corresponding molecular cloud velocity, $V_{\rm cl}$.
The distribution of $V_{\rm up}-V_{\rm cl}$ has a mean value of $-$0.4~km
s$^{-1}$ and a standard deviation of 3.5~km\,s$^{-1}$, much smaller
than the value of 11~\kms\ found using the peak velocity of single-epoch 
\WAT\ maser spectra by Anglada et al. (\cite{ang96}), but similar to the 
3.65~km\,s$^{-1}$ found by Wouterloot et al. (\cite{wou95}).

\noindent
Considering the shape of the upper envelopes (see paper~I and 
Fig.~\ref{upenvs}),
the above implies that the maser emission is maximum for zero projected 
velocities with respect to the local environment. This confirms the well-known 
fact (Elitzur et al.~\cite{eli89}, \cite{eli92}) that the maser emission is 
maximum when the plane of the shocks that create the masing conditions 
is oriented along the line-of-sight.

\noindent
There are a few sources with large values of $|V_{\rm up}-V_{\rm cl}|$ (up to
$\sim$ 7~km\,s$^{-1}$): L1204-G (source number~14) at the negative velocity 
end, and NGC7129/FIRS2 (nr.~12) at the positive end. An offset between the 
velocity of the maser emission and that of the
molecular cloud is however only significant if it is of the order of, or
larger than the width of the maser emission, and if it is persistent in time.
To investigate the former, we show in Fig.~\ref{vhistofrac}a the distribution
of ($V_{\rm up}-V_{\rm cl})/\Delta V_{\rm up}$. $\Delta V_{\rm up}$ is the
second moment of the upper envelope, and as such a measure of the velocity
extent of the maser. For a purely Gaussian distribution, the second moment is
the square of the standard deviation; for the actual shape of the upper
envelopes the precise meaning of $\Delta V_{\rm up}$ is less straightforward,
although it still represents a measure of the velocity dispersion of the 
maser.
Thus, Fig.~\ref{vhistofrac}a shows that the velocity difference between
the maser and the cloud is always less than the width of the maser emission
as measured by $\Delta V_{\rm up}$.
The sources with the largest negative velocity offset are
G32.74$-$0.08 (number~7; ($V_{\rm up}-V_{\rm cl})/\Delta V_{\rm up}=-0.58$),
and L1204-G (nr.~14; $-0.50$); the largest positive offset is for 
NGC7129/FIRS2 (nr.~12; 0.44). 
G59.78+0.06 (nr.~10), in the same bin as nr.~12, has a fractional offset of 
only 0.34. 

\noindent
Fig.~\ref{vhistofrac}b shows the relation of the absolute value of the
difference between the maser- and cloud velocity, relative to the
maser velocity width, as a function of \Lfir. This diagram suggests
that for \Lfir\ $\lsim 7 \times 10^3$~\lsol\ relatively large
values of $|V_{\rm up}-V_{\rm cl}|/\Delta V_{\rm up}$ may occur, while
for the masers pumped by a higher-luminosity YSO the emission that dominates
is at a velocity much closer tothat of the molecular cloud in which they are
embedded ($|V_{\rm up}-V_{\rm cl}|/\Delta V_{\rm up} < 0.2$).

\noindent
To see how persistent these velocity offsets are in time, we have derived for
each source the
``upper envelope'' spectrum separately for the first, middle, and last third
of all spectra, and calculated ($V_{\rm up}-V_{\rm cl}$) in each
case. The resulting distributions are shown in Figs.~\ref{vhisto}b,\,c, and d.
From these histograms we see that with time the three extreme cases identified 
from Fig.~\ref{vhistofrac}a move to the lower bins, i.e.
the difference between the mean maser velocity and the molecular cloud
velocity becomes progressively smaller for all three sources. For L1204-G
(nr.~14) this is because the (more intense) blue-shifted maser lines become
redder with time, while the red-shifted components become bluer (see Paper~I: 
Fig.29); for G32.74$-$0.08 (nr.~7), while the main emission remains
significantly displaced bluewards of the molecular cloud velocity,
red-shifted components appear which shift $V_{\rm up}$ towards $V_{\rm cl}$
(Paper~I: Fig.~15); in NGC7129/FIRS2 (nr.~12; Paper~I: Fig.25) the whole of the
maser emission gradually shifts towards the blue (and towards $V_{\rm cl}$)
with time. 

\begin{figure}[tp]
\resizebox{\hsize}{!}{\rotatebox{270}{
\includegraphics{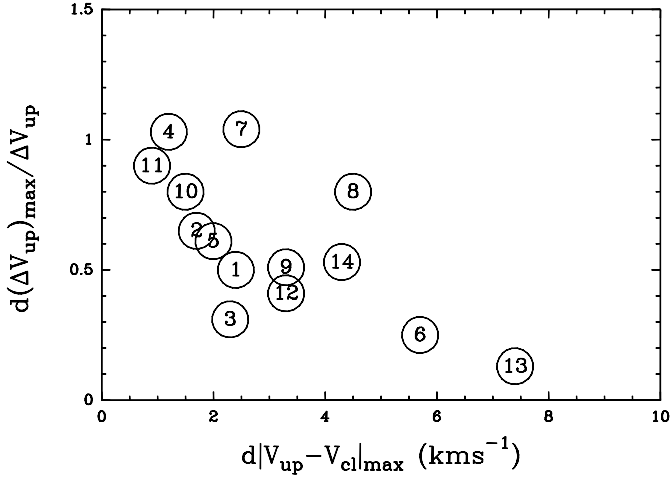}}}
\caption[]{The maximum change in $\Delta V_{\rm up}$ during the three time
intervals 
considered (see text), relative to the $\Delta V_{\rm up}$ derived from
all spectra, as a function of the maximum change in the velocity difference
during those time intervals.}
\label{vmaxchanges}
\end{figure}

\noindent
From a closer look at these data it appears that a variation in the overall 
width of the maser emission is anti-correlated with a variation in its 
velocity.
This is brought out clearly in Fig.~\ref{vmaxchanges}, where
we show the maximum change in $\Delta V_{\rm up}$ during the three time 
intervals considered, relative to the $\Delta V_{\rm up}$ derived from all 
spectra, as a function of the maximum change in the velocity difference during 
those time intervals. Masers that during monitoring have undergone a large 
change in one of those parameters have changed little in the other. 

\noindent
This velocity analysis shows that 
though individual maser components may have 
a large proper motion with respect to the molecular cloud in which they are 
embedded (e.g. Seth et al.~\cite{seth}), the centroid of the maser emission 
remains close to the 
cloud's velocity and, averaged over time, reaches an offset of 
at most half the maser's velocity dispersion as measured by the second moment 
of the ``upper envelope''. There seems to be a critical value of the 
\Lfir\ of the associated YSO, below which larger deviations can occur between
the maser's velocity centroid and that of the molecular cloud than above it.
This result can be understood within the framework of the scenario for maser
emission around a YSO that emerges in the course of this analysis, and which
will become more clear after having considered other observational results
(in particular in Sects.~2.3, 2.4, 2.5, 2.8.3, and 2.8.4).
We assume that around a YSO there are many potential maser sites, that can be
excited by impact with an outflow originating at the YSO if the appropriate
masing conditions can be created. This will be seen to also depend on the
directional properties of the outflow. High-luminosity YSOs may consist of
a collection of lower-luminosity objects, all of which can have an associated
outflow, pointing in different directions, giving rise to a higher degree of
isotropy of the maser emission emanating from the SFR. Lower-luminosity YSOs,
on the other hand, could either be single objects, or less numerous
collections of even lower-luminosity sources. In a SFR with smaller \Lfir,
there will therefore be fewer outflows, and these will be less powerful than
in the high-luminosity SFRs. Fewer maser sites will be excited, and the
emerging maser emission will be more anisotropic in this case. The velocity
at which one detects the maser emission will in this case depend more on
the local morphology of the SFR and on the orientation of the outflows, and
can deviate more from $V_{\rm cl}$ than in the high-luminosity case.

\subsection{Maser luminosity}

\begin{figure}[tp]
\resizebox{\hsize}{!}{\rotatebox{270}{\includegraphics{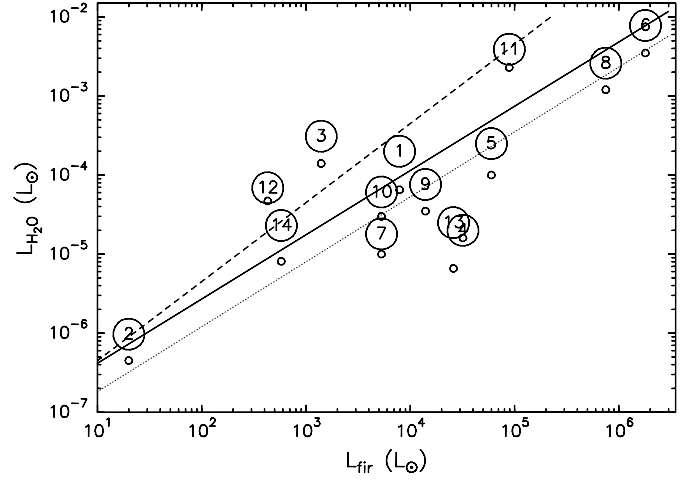}}}
\caption[]{Far-infrared- vs. H$_2$O luminosity for the sample of 14
sources. The drawn line is a least-squares (bisector) fit to the \Lup\ data 
points (large circles; the numbers identify the sources, see 
Table~\ref{tsample}); the dotted line fits the \Lmax\ data (small empty 
circles, each below the corresponding large circle); the dashed line 
represents the fit reported by Wouterloot et al. (\cite{wou95}).}
\label{lum_lum}
\end{figure}

Previous attempts to correlate the maser luminosity with the FIR luminosity
of the associated YSO have always dealt with instantaneous maser spectra 
(Palagi et al.~\cite{PCC93}; Wouterloot et al.~\cite{wou95}).
Given the high variability of the maser emission that we find in almost
all sources, the instantaneous maser luminosity is also highly variable 
(cf. Fig.~\ref{fnrm}).
This may produce two effects: 1) it  might obscure any possible correlation 
existing between the luminosity of the YSO that powers the maser, and the 
maser luminosity, and 2) it might lead one to strongly underestimate the 
conversion factor from YSO luminosity to maser luminosity.

\noindent
For this reason the maximum \WAT\ maser luminosity  \Lup, derived from the 
upper envelope spectra, should represent a more reliable estimate of the 
potential emission of the \WAT\ maser since it eliminates the effects of the 
variability of the individual velocity components.
\Lup\ better approximates the maximum 
maser emission since it gives what the maser would emit {\it if} all the
velocity components were  active {\it at their maximum level} and {\it at the 
same time}.

\noindent
\Lup\  correlates well with the FIR luminosity of the YSO
(Fig.~\ref{lum_lum}). The best fit (drawn line) to the data points 
(encircled numbers) gives
log[\Lup] = ($-7.20 \pm 0.35) + (0.81 \pm 0.07$)log[\Lfir] (corr. coeff.
0.80), i.e.
\Lup\  = $6.37 \times 10^{-8} L_{\rm fir}^{0.81\pm 0.07}$.
This fit agrees reasonably well with that obtained by Wouterloot et
al. (\cite{wou95}; represented by the dashed line: $L_{\rm H_2O}$ = 
4.47$\times 10^{-8} L_{\rm fir}^{1.00\pm 0.07}$), as could be expected since 
our data base contains many observations of a small number of objects, while 
theirs consists of few observations of many objects.
As can be seen from Fig.~7 in Wouterloot et al., at constant \Lfir\ 
the water maser luminosity can change over up to 4 orders of magnitude,
which affects the luminosity derived from instantaneous observations, and may
be explained by changes in the direction of the maser beam and/or
changes in the maser excitation and amplification (see also
Elitzur~\cite{elitzur}). 

\noindent
By its definition, the maser luminosity derived from the upper envelope 
represents an upper limit to the 
instantaneous maser luminosity. In Fig.~\ref{lum_lum} we
therefore also show the fit to \Lmax, which is derived from the
actually observed maximum integrated flux density. This fit is described by
\Lmax\  = $2.75 \times 10^{-8} L_{\rm fir}^{0.82\pm 0.08}$
(corr. coeff. 0.79), and is shown as the dotted line.
As expected, it lies below the relation defined by \Lup\ (by $\sim 0.4$ in
log[L$_{H_2O}$]).
We find \Lmax /\Lup\ 
to be between 0.25 (Sh~2-269~IRS2; source nr. 5) and 0.80 (Mon~R2~IRS3;
nr. 4); there is no correlation between this ratio and \Lfir.

\subsection{A variability index for the maser emission}

\begin{figure}[tp]
\resizebox{\hsize}{!}{\rotatebox{270}{\includegraphics{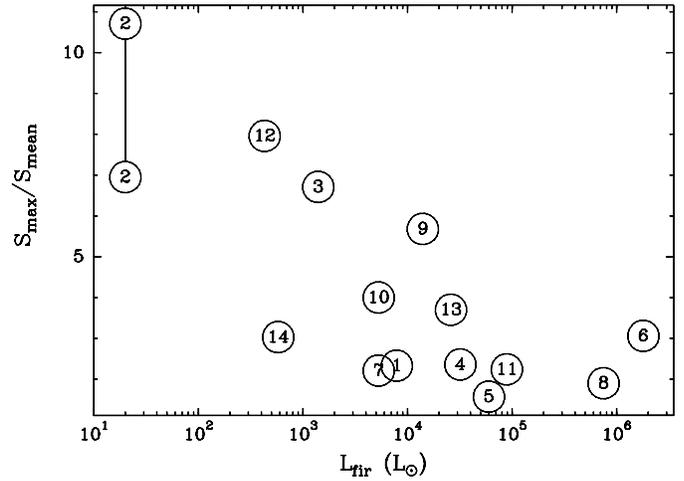}}}
\caption[]{Ratio between the maximum integrated flux density ever observed, 
$S_{\rm max}$, and the mean value of the integrated flux density, 
$S_{\rm mean}$, over the whole monitoring period, as a function of the YSO 
FIR luminosity. The two appearances of object nr.~2 (L1455~IRS1) represent 
different ways to take into account the numerous non-detections of the maser
(see text).}
\label{ff_lum}
\end{figure}

The definition of a variability index to describe with a single parameter
the variability of a maser (or even a single emission component) is an almost
impossible enterprise in view of the complex patterns shown in the
velocity-time-intensity plots (paper~I and Fig.~\ref{grey}).
With these limitations, and aiming to capture the overall variation of the
maser emission, we have derived for each source the ratio
$S_{\rm max}$/$S_{\rm mean}$ between the maximum and the mean integrated flux
densities over the whole monitoring period.
An anti-correlation is found between the YSO FIR luminosity and this ratio
(Fig.~\ref{ff_lum}). Clearly, {\it high-luminosity sources tend to be
associated with more stable masers, while lower luminosity ones have a more
variable emission}. Note that L1455~IRS1 (source nr.~2), the maser emission of
which often disappears below our detection limit, appears twice in
Fig.~\ref{ff_lum}, because $S_{\rm mean}$ was calculated in two ways: by
assigning to the non-detections either a value of 0 (resulting in the higher
value of the plotted ratio) or $3 \sigma (V_{\rm max} - V_{\rm min})$
(resulting in the lower value of the plotted ratio), where $V_{\rm max}$ and
$V_{\rm min}$ are the extreme velocities of the corresponding upper envelope
spectrum, and $\sigma$ is the rms-noise in an individual maser spectrum.

\noindent
In lower-luminosity YSOs a smaller number of maser components gets excited 
(see also Sect.~2.5) and their intrinsic time-variability will affect the 
total output more than in higher-luminosity YSOs, where a much larger number
of components might be simultaneously excited, thus reducing the effect of 
their individual time-variability on the total maser output. Moreover,
for increasingly smaller \Lfir, the conditions of maser amplification are more
likely to be closer to the threshold conditions and consequently the maser
emission will be more unstable.

\subsection{Velocity range of the maser emission}

In order to characterize the velocity  width of the emission we use the second 
moment of the ``upper envelope'' spectrum, $\Delta V_{\rm up}$. 

\begin{figure}[tp]
\resizebox{\hsize}{!}{\rotatebox{270}{\includegraphics{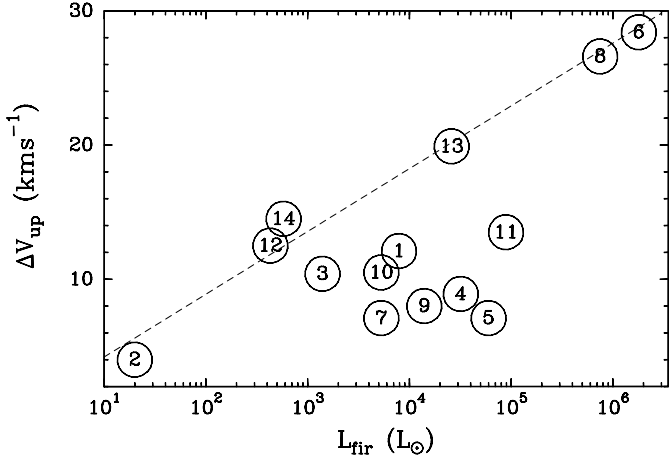}}}
\caption[]{Plot of the second moment of the ``upper envelope'' spectrum
as a function of the YSO FIR luminosity. The dashed line is a (bisector)
least-squares fit through 6 data points defining an upper boundary to the data:
$\Delta V_{\rm up} \propto L_{\rm fir}^{4.68 \pm 0.26}$.
}
\label{dv_lum}
\end{figure}

\noindent
$\Delta V_{\rm up}$ is shown as a function of the
YSO FIR luminosity in Fig.~\ref{dv_lum}. It seems that high values of
$\Delta V_{\rm up}$ are encountered only for high \Lfir, while lower
$\Delta V_{\rm up}$ values occur at any FIR luminosity. Thus,
more luminous sources {\it can} excite maser emission over a larger velocity
interval, but apparently do not necessarily always do so.
Therefore, what the upper envelope shows is {\sl potential}. Rather than
showing a relation between the data points, Fig.~\ref{dv_lum} defines an upper
boundary indicating that there is a maximum {\sl possible} velocity extent of the
maser emission,
which depends on the FIR luminosity of the associated (pumping-) source. This
upper limit is shown in Fig.~\ref{dv_lum} as a dashed line, defined by the
sources 2, 6, 8, 12, 13, and 14:
$\Delta V_{\rm up} \propto L_{\rm fir}^{4.68 \pm 0.26}$.

\noindent
Felli et al. (\cite{FPT92}) found a correlation  between the maser
luminosity and the mechanical luminosity of the associated molecular
outflow (see also Lada~\cite{lada85}), supporting the hypothesis that
maser conditions are created
where the molecular outflows shock the surrounding molecular gas.
The mechanical luminosity of the outflows is also correlated
with the FIR luminosity of the YSOs. Similarly, Wouterloot et al. 
(\cite{wou95}) found that for \Lfir $> 10^2$~\lsol\ IRAS sources with 
``maser-like'' colours have significantly larger CO (FWHM) linewidths than 
those with ``non-maser-like'' colours, and that
this is even independent of whether a maser has actually been detected or
not. The CO (FWHM) linewidth was found to depend only weakly on
\Lfir: an increase of a factor of about 2 between \Lfir=$10^2$ and $10^6$.

\begin{figure} 
\resizebox{\hsize}{!}{\rotatebox{270}{\includegraphics{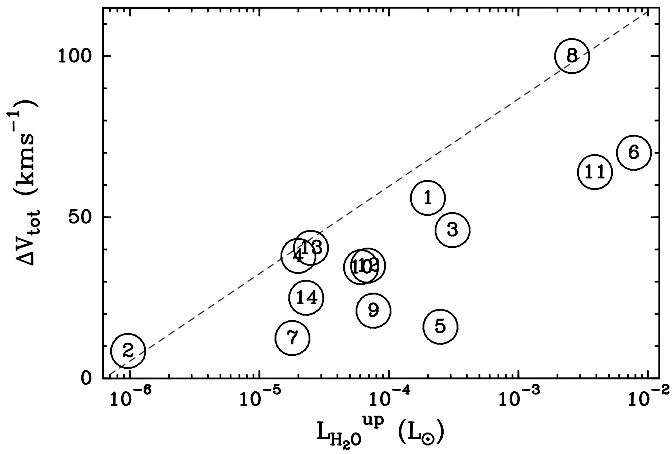}}}
\caption[]{Total velocity range of the maser emission, $\Delta V_{\rm tot}$
determined from the
frequency-of-occurrence histograms, as a function of \Lup. The dashed line is 
a (bisector) least-squares fit through 4 data points defining an upper 
boundary to the data: 
$\Delta V_{\rm tot}$ = (168.1 $\pm$ 4.5) + (27.12 $\pm$ 1.08)log\Lup.}
\label{vtotlup}
\end{figure}

\vskip 0.3cm
\noindent
The use of $\Delta V_{\rm up}$ to characterize the velocity range of the maser
emission can be misleading, however:
if the maser emission is dominated by one or a few strong components, then
$\Delta V_{\rm up}$ will be small, even though there may be many weaker
components with a large range of velocities. 
These weaker outlying components are taken into account if the velocity
range of the maser emission ($\Delta V_{\rm tot}$) is taken to be the total 
velocity extent in the frequency-of-occurrence histograms (see Paper~I and 
Fig.~\ref{freq}). The general trend is that $\Delta V_{\rm tot}$ increases 
with \Lup, \Lmax, and \Lfir. As an example we show this for \Lup\ in 
Fig.~\ref{vtotlup}.

\noindent
Just like Fig.~\ref{dv_lum}, Fig.~\ref{vtotlup} shows that while in stronger 
maser sources (associated with more luminous YSOs, see Fig.~\ref{lum_lum}) 
emission can be excited over a larger range in velocity, there is no 
guarantee that this will be so: many of the more luminous objects have a 
relatively small $\Delta V_{\rm tot}$. An example is Sh~2-269~IRS2 (source 
nr. 5),
where the maser emission, while strong, is always contained within a rather
narrow velocity interval (at least at the 5$\sigma$-level used to construct 
the frequency-of-occurrence histograms). A different case is Mon~R2~IRS3 
(nr. 4), a strong maser with emission virtually always in a very narrow range 
of velocities (see Paper~I, Fig.~9), but which during our long monitoring 
campaign occasionally showed components both blue- and red-shifted by up to 
20~\kms, thus increasing $\Delta V_{\rm tot}$.
Thus, also in Fig.~\ref{vtotlup} the data points define an upper 
boundary indicating the maximum {\sl possible} velocity range of the maser 
emission as a function of \Lup. In Fig.~\ref{vtotlup} this envelope is 
determined by sources 2, 4, 8, and 13. A bisector fit through these data 
points gives $\Delta V_{\rm tot}$ = (168.1 $\pm$ 4.5) + 
(27.12 $\pm$ 1.08)log\Lup, which is shown 
as a dashed line. Note that the value of the intercept is influenced by the
sensitivity of the data, by the 5$\sigma$ detection limit adopted in the
construction of the frequency-of-occurrence histograms, as well as by the 
fact that there may be emission at velocities not within our frequency band 
(see the velocity-time-intensity diagrams in Paper~I).

\subsection{Distribution of velocity components}

Our single-dish observations can resolve the maser emission only in the
velocity domain, but not in the spatial domain. Interferometric observations 
(see e.g. VLBA observations by Seth et al.~\cite{seth}) have shown that a
single velocity component may arise from as many as a dozen spatially
separated maser spots. With the provision that the number of spectral 
components observed in our spectra does not necessarily correspond directly to 
the number of individual maser components present in the SFR, by studying
their distribution in velocity we can still derive useful information on how
the energy input from one or more YSOs in a SFR is distributed in the
outcoming maser spectrum, and how the situation is affected by the
luminosity of the YSOs, or, ultimately, how the dynamics of the molecular
cloud surrounding the YSOs is affected by their presence.

\begin{figure}[tp]
\resizebox{\hsize}{!}{\rotatebox{270}{\includegraphics{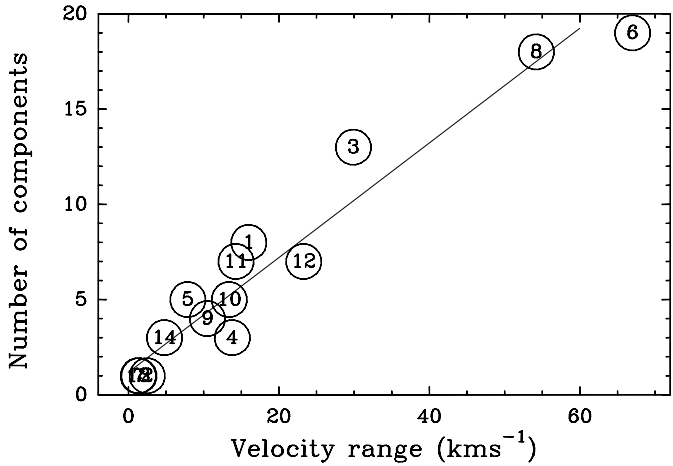}}}
\caption[]{The number of emission components, $Nc$, in the spectrum with the 
highest integrated flux density, as a function of the velocity range of the 
emission, $Vrange$, in that spectrum. The drawn line is a (bisector) 
least-squares fit to all data points: 
$Nc = (1.2 \pm 0.4) + (0.30 \pm 0.02)~Vrange$}
\label{ncompvrange}
\end{figure}

\smallskip\noindent
To study this effect we used the spectrum with the highest integrated flux 
density during the monitoring period, in which we have counted the number of
individually visible spectral components, including emission down to levels
of $2-3\sigma$ (which however varies from spectrum to spectrum).
(This is practically impossible to do from either the upper envelope spectrum 
or the frequency-of-occurrence histograms, where the individual
components appear indistinguishably merged due to (random and 
systematic) velocity shifts during the monitoring period.)  
In Fig.~\ref{ncompvrange} we plot for each source the number of components in 
the spectrum with the maximum integrated flux density as a function of the 
total velocity range of the emission in that spectrum. Note that both 
quantities plotted are {\it lower limits}, due to a combination of 
sensitivity, spectral 
resolution, and the intensity of individual components. Nevertheless, there is 
a clear and rather tight correlation between the number of components in a 
spectrum and the velocity range over which they are found.
A bisector fit through all data points gives the number of components, $Nc$, 
as a function of velocity range, $Vrange$: 
$Nc = (1.2 \pm 0.4) + (0.30 \pm 0.02)~Vrange$ (corr. coeff. 0.97). One way
of looking at this relation is that for every 10~\kms\ increase in $Vrange$, 
three more components are found.

\begin{figure*}[tp]
\resizebox{15cm}{!}{\rotatebox{270}{
\includegraphics{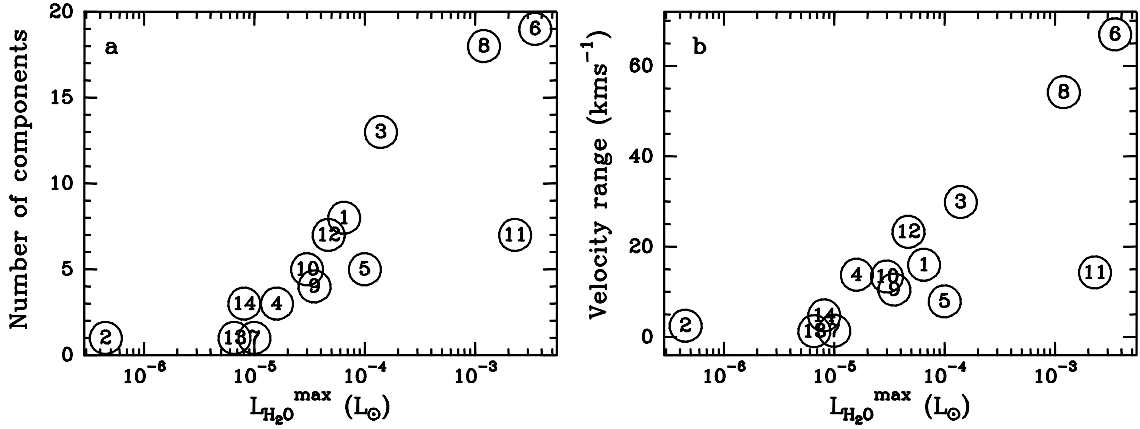}}}
\caption[]{Correlation between the number of components ({\bf a}) and the
velocity range of the maser emission ({\bf b}) in the spectrum with
the highest integrated flux density, as a function of \Lmax, the maser
luminosity determined from that same spectrum. Both quantities increase with
increasing maser luminosity.
}
\label{ncvlmax}
\end{figure*}

\noindent
The number of components and the velocity range are plotted separately as a 
function of \Lmax\ in Fig.~\ref{ncvlmax}a and b, respectively. We see that 
both quantities increase with increasing maser luminosity. 

\noindent
Note that in Fig.~\ref{ncvlmax}b we find a correlation between 
$Vrange$ and \Lmax, rather than just an upper boundary, as was the case
for \Lup. This is because \Lmax\ is {\it directly related} to the velocity
range, in the sense that it pertains to the same spectrum, while \Lup, which
measures some fictitious maximum maser luminosity, can be anywhere between
1.25 and 4 times larger than \Lmax\ (see Sect.~2.2).

\noindent
Figures~\ref{ncompvrange} and \ref{ncvlmax} show that {\it higher maser power 
goes into more emission channels, that are spread over a larger range in 
velocity}. Both the number of components and the velocity range of the 
emission seem to be insensitive to \Lfir\ (not shown): higher
\Lfir\ allows a larger velocity extent of the maser emission, but does not
impose it. As a consequence, and perhaps surprisingly, the maximum velocity
range (and the largest number of components) are not necessarily reached in
the spectrum with the largest integrated flux density. 

\noindent
In a SFR there are likely to be many potential sites of maser amplification, 
which can be excited if there is sufficient energy to pump them. Water masers 
are excited behind shocks (Elitzur et al.~\cite{eli89}), which are likely to 
be caused by outflows or jets driven by the associated YSO 
(Felli et al.~\cite{FPT92}).
To excite maser emission at large (relative to the ambient molecular cloud) 
velocities requires sufficiently powerful jets and outflows from the YSO to 
provide the necessary 
energy. Hence, the more luminous YSO's will be associated with maser spectra
containing more emission components over a larger range in velocity, as we
indeed find. The basic pumping process seems to be the same regardless of the 
YSO luminosity, but in SFRs with lower input energies (i.e. a lower 
YSO-luminosity driving a lower-velocity outflow) only components with low
(relative to the ambient molecular cloud) velocities can be excited. 

\subsection{Velocity drifts}

In all sources, most if not all spectral features undergo velocity
drifts, and a number of them have been identified in Paper~I. 
These can be recognized in the large scale velocity-time-intensity plots of 
Paper~I as inclined linear structures, indicating systematic changes of 
line-of-sight velocity of the masing gas with time.

\noindent
Indirect evidence that velocity drifts must be observable in maser 
components comes also from VLBI observations (e.g. Seth et al.~\cite{seth}). 
From
studies of the spatial distribution and proper motions of maser spots, three 
type of maser components are found: 1) in a rotating disk around the YSO; 
2) in a high-velocity collimated bipolar outflow originating from the YSO and 
perpendicular to the disk; and 3) at the bow shocks produced by the outflows. 
Velocity drifts can occur due to rotation of the disk, to acceleration or
decelaration in the 
collimated outflow, or to precession of the jet/outflow. One does not need 
high spatial resolution to study these velocity drifts, as they  
are observable also with single-dish monitoring of the type 
presented in this work (e.g. Cesaroni~\cite{cesa}; 
Lekht et al.~\cite{lekht140}).

\noindent
With a spectrum every 2--3
months (i.e. once every $\sim 60-100$~days), and a spectral resolution of
$\sim 0.16$~\kms, the minimum detectable velocity drift from spectrum
to spectrum is 
$\gsim 0.6-1.0$~\kmsyr. However, if a component can 
be traced over a longer period of time, then the minimum detectable value can 
be much less than this.
In fact, for the 15 emission components that we have analyzed 
we find velocity gradients between 0.02 and 1.8~\kmsyr. 
The lower value is found for components that could be traced over the whole 
$\sim 4600$~day period of monitoring, while the higher value was found for a 
burst-component with a duration of $\sim 63$~days. Considering the small
number of components studied in detail, we find equal numbers of negative
(9/15) and positive (6/15) velocity gradients.

\noindent
Since the intensity of the velocity component during the drift is far from 
constant, one might object that what we see is not due to a unique component 
drifting in velocity, but rather to a casual sequence of 
small bursts each one occuring at a slightly different velocity and with the 
proper time delay with respect to the preceeding one, as the short
time-duration of the spatial-velocity components found by Seth et
al. (\cite{seth}) from VLBA observations might suggest. Obviously, we have no 
means to reject this second explanation and we believe that in the spectra of 
sources which have a large number ($>$10) of velocity components bursting in a 
random fashion it is impossible to make any statement of this type. Another 
difficulty arises from the fact that many features have very short life-times 
and in such small time interval ($\sim$ one year) the effect of a velocity 
drift may be less evident.

\begin{figure}
\resizebox{\hsize}{!}{\includegraphics{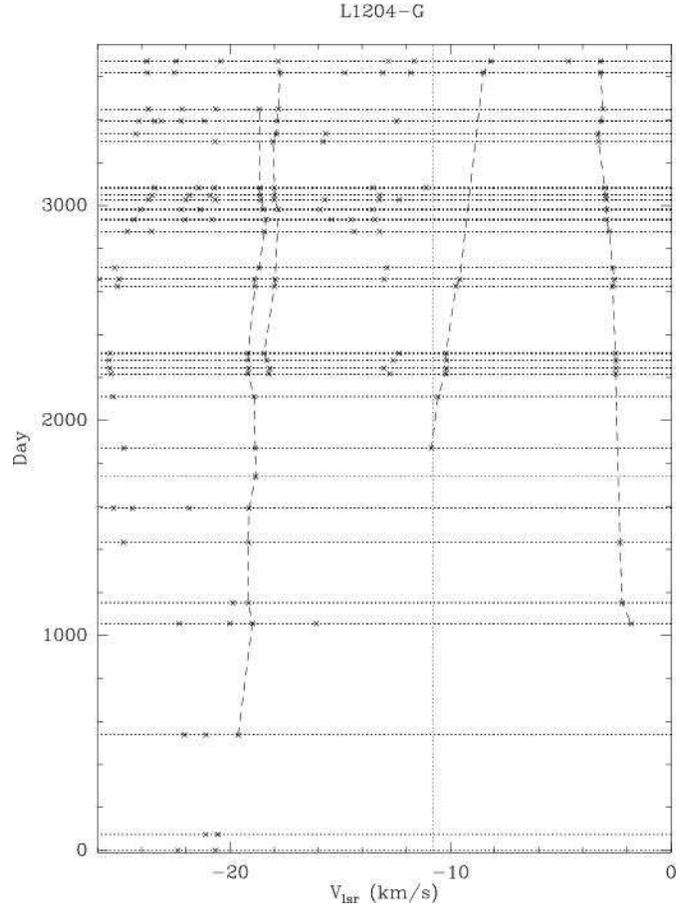}}
\caption[]{Overview of the velocity of the emission components in L1204-G
spectra, derived from Gaussian fits. The horizontal
dotted lines indicate the days (averaged in bins of 10~days) for which spectra
are available; the vertical dotted line indicates the velocity of the
surrounding high-density molecular gas ($V_{\rm cl}$). The crosses
mark the central velocity of each fitted emission peak;
Velocities considered to identify single components are
connected by a dashed line, if they are visible over a considerable amount
of time.}
\label{l1204goverview}
\end{figure}

\noindent
This is why we have limited ourselves to the most obvious cases, in particular
to sources with few velocity components, to sources in which one component
dominates the spectrum, or to sources with components at velocities
far from the more crowded part of the spectrum.
One of the best examples is L1204-G (Fig.~\ref{l1204goverview}) where at least
four components (two around $-$19~\kms, one near $-$10~\kms, and one at 
$-$2.5~\kms) are seen drifting in opposite directions almost throughout the 
entire monitoring period. For the
$-$2.5~\kms\ component we derive a velocity drift of $-0.10$~\kmsyr\ for the
first $\sim 1500$~days after its first detection, followed by a more rapid
deceleration of $-0.40$~\kmsyr\ in $\sim 680$~days (associated with an
increase of this component's flux density from 2 to 20~Jy). The $-$19~\kms\
components changes velocity in a more erratic way: while the general
trend is for the velocity to become redder, there are periods in which the
velocity is constant or becomes bluer. The $-$10~\kms\ component shows a
velocity gradient of $+1.23$~\kms\ during the first $\sim 790$~days after its
first appearance around day 1850 ($+0.57$~\kmsyr); after reaching a peak in
flux density of $\sim 50$~Jy on day 2243 it rapidly dropped below our
detection limit, only to resurface near the end of the monitoring period (in
the last spectrum $F \approx 97$~Jy), with a velocity that agrees with an
extrapolation of the gradient found from its previous appearance.
In Fig.~\ref{l1204goverview} we also
see a component at \vlsr $\simeq -24$~\kms, near day=3000. In the 6
observations taken around that date (day 2879$-$3083), the line exhibits a 
redshift of 0.7~\kmsyr; including also the observation at day=2711, the 
redshift amounts to 1.8~\kmsyr. The velocity-change coincides with a period 
of decline after an increase in flux density.

\noindent
A behaviour similar to that found for the $-2.5$~\kms\ component in L1204-G
has been found in W75N by Hunter et al. (\cite{hun94}) who interpreted the
drift as an outward acceleration. Likewise, the reddening of the components
at $-19$ and $-10$~\kms\ could be taken as an outward acceleration of maser 
features on the far side of the driving source. 
However, as summarized at the beginning of this sub-section, other 
explanations are possible, including precession of
the jet exciting the maser, and rotation of the maser component
around the YSO,
as well as random superposition of short bursts of spatially separated
components at similar velocities, even though it seems improbable that this
explanation might work for steady drifts persisting over many years. Only a
finer time-sampling of single-dish observations together with frequent VLBI
observations can clarify this issue.

\subsection{Maser bursts: duration, intensity and linewidth} 

A burst is a rapid increase of the flux density of the maser
emission at a given velocity $V$ during a certain (usually brief) time.
The change in flux density is defined as
$\Delta F(V) = [F_{\rm peak}(V)-F_{\rm steady}(V)]$, where $F_{\rm steady}(V)$
is the (average of the) intensity level immediately before and after the
burst at velocity $V$. During the burst, the flux density reaches a maximum 
$F_{\rm peak}(V)$.
The burst's duration is $\Delta t(V) = [t_{\rm final}(V)-t_{\rm initial}(V)]$,
where $t_{\rm initial}(V)$ and $t_{\rm final}(V)$ denote the time of start
and end of the flux density increase, respectively. 

\noindent
Given the variability of the emission and the relatively long time between
two consecutive observations, it is often difficult to determine these
parameters,
and the determination of $\Delta F(V)$ and $\Delta t(V)$ is not homogeneous
and rather subjective.
The ever-present intrinsic variability can make it difficult to distinguish
a burst, unless the increase in flux density is particularly large with
respect to the `normal' variability.
$\Delta t(V)$ depends on the rapidity of the flux
density change, shorter bursts being easier to isolate. 
Finally, a proper determination of $\Delta t(V)$ is influenced by the uneven 
time coverage of our monitoring, and by the minumum sampling interval.
Hence, $\Delta F(V)$ tends to be a lower limit, $\Delta t(V)$ an upper limit.

\noindent
We have not done any systematic analysis of bursts, but as with the
analysis of velocity gradients in the previous sub-section we have selected
a few of the more evident examples: we selected 14 bursts in 9 emission
components, in 6 sources. In this small sample we find increases of flux
density ($\Delta F(V)$) from about 40\% to $\gsim 1840$\% with respect to 
the $F_{\rm steady}(V)$-level;
the largest absolute flux density increase was found in Mon~R2 
($\sim 820$~Jy [75\%], in a burst lasting 63 days).
$\Delta t(V)$ ranges from a minimum value equal to our sampling interval
($\sim 60$~days) to up to $\sim 900$~days, mainly because variations of
longer duration are not defined as bursts. However, some type of
long-term (`super-') variability is visible in the behaviour of 
$\int F{\rm d}v$ of several sources: the integrated flux density 
of all maser components changes more or less regularly with time. 
We will discuss this long-term variability in Sect.~2.8.5.

\smallskip\noindent
For W49N, Liljestr\"om \& Gwinn (\cite{lil00}; their
Fig.~10) find that the duration and intensity of the maser 
outbursts depend on the
velocity offset with respect to $V_{\rm up}$: high-velocity blue-shifted and
red-shifted components seem to have shorter duration and smaller flux density
increase than those close to $V_{\rm up}$.
We do see an indication for this from the shapes of
the upper envelopes (see Paper~I and Fig.~\ref{upenvs}) of our
sources, in which the flux density tends to decrease very rapidly at
velocities away from $V_{\rm up}$. At the same time
the frequency-of-occurrence diagrams (Paper~I and Fig~\ref{freq}) show that
maser components at large blue- and red-shifted velocities (with respect to
$V_{\rm fr} \approx V_{\rm up}$) are detected only a fraction of the time, 
compared to the components near $V_{\rm up}$, implying that the {\it lifetime} 
of these maser components is shorter.
It should be noted though, that weaker, short-duration components that occur 
in the central, most crowded part of the spectra, are virtually impossible to 
identify. 

\noindent
Clearly, the two
dependencies call for a unique explanation. The geometrical one seems
to be the simplest and most widely accepted: if masers occur in shocks
and the peak of the maser-beaming is in the plane of the shock, then the
both the maximum emission and the largest $\Delta F(V)$ will be observed 
from planes closely aligned with the line-of-sight, and at a line-of-sight 
velocity near the systemic velocity ($V_{\rm cl}$). 
Small changes in the amplification of the maser will produce larger 
absolute changes in flux density.
At velocities near $V_{\rm cl}$ we are likely to see the cumulative effects
of more than one maser spot, as all spots with the plane of the shock along
the line-of-sight 
will be in the condition of maximum beaming.

\smallskip\noindent
Liljestr\"om \& Gwinn (\cite{lil00}) suggest that the duration of the burst
is determined by the time required by the shock to propagate across the
maser filament, $\Delta t(V)$ = D/V, where D is the transverse diameter
of the maser. Assuming a mean value of D$\sim$1~AU and V$\lsim$55~\kms, which
in our sample is the maximum velocity offset between a maser component and 
the velocity of the cloud in which it is embedded, we obtain
$\Delta t(V) \gsim 32$~days, which is at the limit of our sampling rate.

\noindent
This would also explain the dependence of $\Delta t(V)$
on the velocity offset. The suggestion of Liljestr\"om \& Gwinn (\cite{lil00})
is that high-velocity offsets from $V_{\rm up}$
select higher velocities in the outflow and hence higher shock 
velocities and smaller $\Delta t(V)$. 

\noindent
Another quantity that has often been discussed in relation to maser bursts
is the change of the linewidth during a burst. 
In the (small) number of bursts investigated by us, we do not see any 
systematic behaviour in $\Delta V_{\rm fwhm}$ with 
flux density, which agrees with the Liljestr\"om \& Gwinn (\cite{lil00})
study of bursts in W49N.

\subsection{Comparison of overall maser properties as a function of FIR 
luminosity}

\begin{figure}
\resizebox{\hsize}{!}{\includegraphics{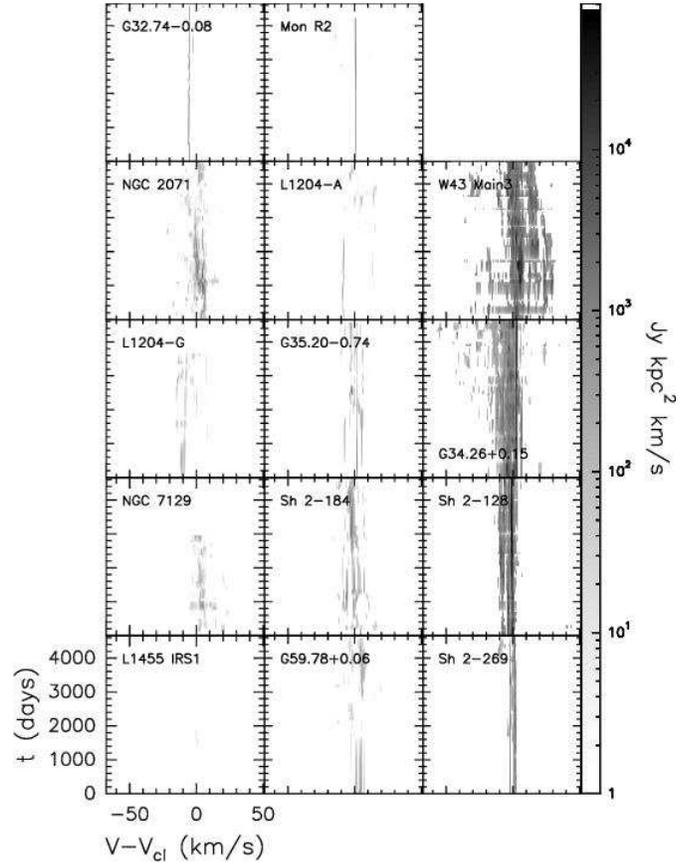}}
\caption[]{Grey-scale maps of the flux density, multiplied by distance$^2$; cf.
Paper~I. The velocity scale is relative to the cloud velocity $V_{\rm cl}$.
\Lfir\ increases from bottom to top and from left to right; Mon~R2~IRS3 has
\Lfir $= 3.2 \times 10^4$~\lsol.}
\label{grey}
\end{figure}

\subsubsection{The scaled velocity-time-intensity plots}

In Fig.~\ref{grey} we show the velocity-time-intensity plots in order of
increasing FIR luminosity (\Lfir\ increases from bottom to top, and left to 
right). 
The velocity-scale is the same in all plots and the velocities are referred 
to  $V_{\rm cl}$ (i.e. the quantity on the horizontal axis is 
$V_{\rm observed} - V_{\rm cl}$).
The intensity in all plots has been normalized to the same
distance, by multiplying the values by (d[kpc])$^2$.
The (logaritmic) intensity scale in all plots is the same so that 
Fig.~\ref{grey}
shows what one would see if all the sources were at the same distance.

\noindent
Note that for \Lfir $\gsim 3 \times 10^4$~\lsol\ the emission becomes
increasingly complex: going from Mon~R2~IRS3 (\Lfir $= 3.2\times 10^4$~\lsol)
to W43~Main3 (\Lfir $= 1.8\times 10^6$~\lsol) the maser emission changes from
being dominated by a single component to being highly structured and 
multi-component; the
velocity extent of the emission also increases. For \Lfir $\lsim 3 \times
10^4$~\lsol\ on the other hand, while the maser emission shows the same
variety of morphologies, from the single-/dominant component-type to a
(modest) degree of complexity, there is no systematic trend with \Lfir\ 
and the velocity extent of the maser emission remains smaller than what 
is found for the highest-luminosity sources (cf. Fig.~\ref{vtotlup}, 
Sect.~2.4).
The source with the lowest \Lfir\ (20~\lsol; L1455~IRS1: nr. 2) is again a 
special case, where for much of the time the maser has not been detected.
This can be understood by the explanation given for the variability in
Sect.~2.3, and 
is consistent with the results of a 13-month maser-monitoring program 
of low-luminosity YSO's by Claussen et al. (\cite{cla96}): below a YSO 
luminosity \Lfir $\sim$25 L$_\odot$ there is a higher degree of variability 
of the maser emission, which often disappears below the detection threshold.

\begin{figure}
\resizebox{\hsize}{!}{\includegraphics{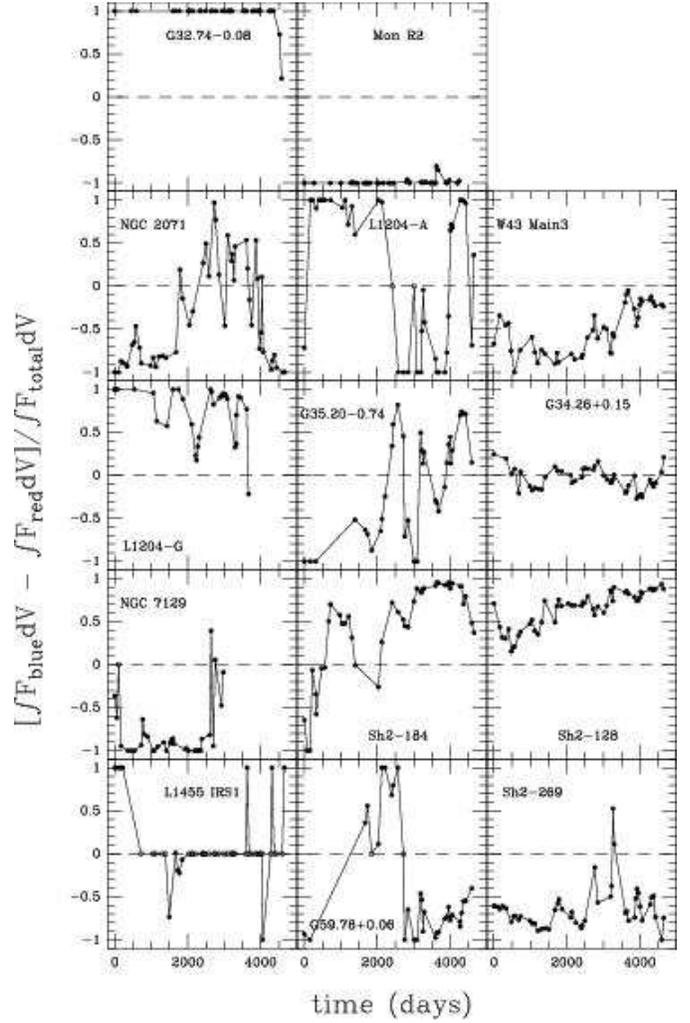}}
\caption[]{The normalized difference between the integrated (Jy\, \kms) blue
and red parts of the spectra of all sources, as a function of time.
Integration was performed from $- \infty$ to $V_{\rm cl}$ (blue) and from
$V_{\rm cl}$ to $\infty$ (red), where the reference velocity $V_{\rm cl}$ is
that of the molecular cloud (see Table~\ref{tsample}). Open circles indicate
cases where both integrals are zero (i.e. no maser detection).
The panels are ordered in \Lfir, which increases from bottom to top and from
left to right.}
\label{redblue}
\end{figure}

\subsubsection{Red and blue asymmetries in the spectra}

In Fig.~\ref{redblue} we show the relative strengths of the integrated blue
(integrating between $- \infty$ and $V_{\rm cl}$) and red ($V_{\rm cl}$ to
$\infty$) maser emission as a function of time. The difference between the 
blue and red integrals is normalized by the total integrated flux.
The panels are ordered in \Lfir\ as in Fig~\ref{grey}.
In this figure, masers with predominantly blue (red)-shifted emission with
respect to $V_{\rm cl}$ have positive (negative) values along the ordinate;
masers with {\it only} blue (red) emission have ordinate values of +1 ($-$1).
Interpretation of this diagram is not straightforward, as changes in the
relative strengths of the blue and red sides of the spectra
may have various origins: the intrinsic variability of all maser components
(all sources); velocity drifts of the emission with respect to $V_{\rm cl}$
(e.g. L1204-G, NGC~7129); the sudden appearance of strong components at a
velocity on the opposite side of $V_{\rm cl}$ with respect to the bulk of the
emission (e.g. G32.74$-$0.08, Mon~R2); the sudden {\it dis}appearance or
weakening of a component (e.g. Sh~2-269, where the smaller of the two peaks,
near day 2800, coincides with an abrupt weakening of the red-shifted
strong 19.5~\kms\ and (the weaker) 20.8~\kms\ components, while the
blue-shifted 16.2~\kms\ remains the same, thus causing a relative increase
of the blue-shifted integrated emission); flaring of individual components
(e.g. Mon~R2, Sh~2-269; in this latter source the strong peak showing a
temporary dominance of the blue part of the maser spectrum is caused by a
burst in the 16.2~\kms\ component near day 3200, which has no counterpart in
the other main (and red-shifted) components, thus causing the temporary
dominance of the blue-shifted emission seen in this panel); or combinations
of the above. One can only fully disentangle all the information contained
in these diagrams when it is used in combination with the 
velocity-time-intensity diagrams, and with diagrams showing the behaviour of 
the first moments (velocity) of the blue- and red-shifted sides of the maser
spectra.

\noindent
While there are several maser sources that emit almost exclusively on the
blue (L1204-G, G32.74$-$0.08, Sh~2-128) or red (NGC~7129, Mon~R2, Sh~2-269,
W43~Main3) side, in about half of the objects in our sample the emission shows
no dominant side. In some sources the dominant emission switches frequently
between the blue and red sides (e.g. G34.26+0.15, G35.20$-$0.74), while in
other objects there seem to be longer periods in which one side of the
spectrum prevales over the other (e.g. NGC~2071, G59.78+0.06, L1204-A).
We do not see any obvious systematic dependencies on \Lfir\ from this figure.

\begin{figure}
\resizebox{\hsize}{!}{\includegraphics{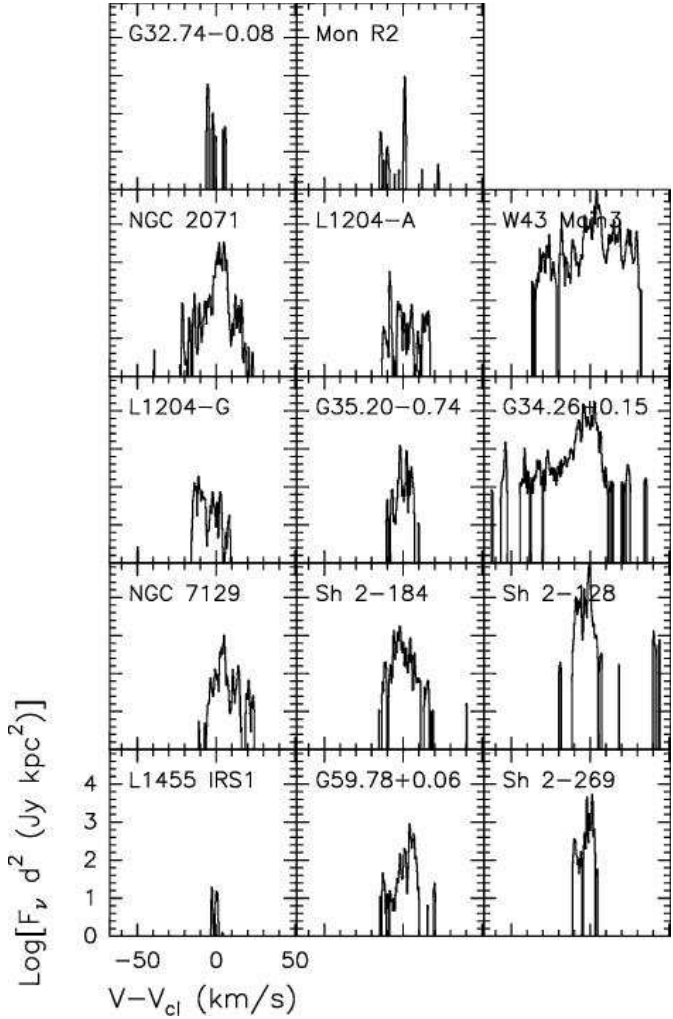}}
\caption[]{Upper envelopes of all sources, scaled by dist$^2$.
The panels are ordered in \Lfir, which increases from bottom to
top and from left to right.}
\label{upenvs}
\end{figure}

\subsubsection{The scaled upper and lower envelopes}

\noindent
There are two ways to study the minimum level of activity of the maser.
The first is through the light curve, which is obtained by integrating the
spectra in velocity, and which will be considered
in Sect.~2.8.5, the second is via the lower envelope which is a function
of the velocity but time-independent(see Sect.~1 for the definition of
the lower envelope). Obviously, the
two ways may give different answers. For instance the lower envelope can
be zero over the entire velocity range because of short-duration
random bursting at different non-overlapping velocities and, at the same
time, the integrated flux may be constantly above zero.
Considering sources of increasing FIR luminosity (see ordering in
Table~\ref{tsample}), the lower envelope is zero
in the first four sources; there is a small peak of $\sim$ 15\% of the
maximum for the fifth source G32.74$-$0.08, which
is peculiar because it emits essentially at only one velocity.
Then there are other four sources with a zero lower envelope,
and finally the last five sources (all with \Lfir $> 3\times 10^4$~\lsol)
show a small peak, always at (within 1 resolution channel of) the velocity
of the main peak in the upper envelope (and close to that of the dense 
molecular gas, $V_{\rm cl}$) and with 
an intensity of 1$-$20\% relative to it. Since at these percentages
of the peak emission all the sources in our sample are above the
noise level of our observations, this does not seem to be an
sensitivity effect. We conclude that only SFRs with high FIR luminosity
($> 10^4$~\lsol) are capable to maintain a certain level of emission at a
given velocity (basically the velocity of the peak of the upper envelope 
spectrum) for extended periods of time (see also Sect.~2.8.4). 
Considering that at $V_{\rm cl}$ 
the velocity component along the line-of-sight is zero, the above result 
{\it does not} imply that the {\it same spot} will always be active, since 
all spots with the plane of the shock along the line-of-sight (which are also 
the most intense ones) are indistinguishable in our observations (see also 
Sect.~2.7). 

In Fig.~\ref{upenvs} we show the upper envelopes in order of
increasing FIR luminosity (cf. Figs.~\ref{grey} and \ref{redblue}).
It appears that sources with
$4 \times 10^2$~\lsol $\lsim$ \Lfir $\lsim  6 \times 10^4$~\lsol\ (from
NGC7129 to Sh~2-269 in order of increasing luminosity) have similar upper
envelopes, with values of log$[F_{\nu} d^2] \sim 3$ and with comparable
velocity range. Outside this homogeneous group of sources we find on one 
hand the
lowest-luminosity source (\Lfir $\approx 20$~\lsol), with log$[F_{\nu} d^2]$
two orders of magnitude smaller and with a narrower velocity range of the
emission, and on the other hand the sources with higher luminosities 
(\Lfir $\gsim 6 \times
10^4$~\lsol), with peak values of log$[F_{\nu} d^2] \sim 4-5$ and a larger
extent in velocity. These distinctions may reflect three different regimes of
maser excitation:

\begin{enumerate}
\item In the lowest luminosity sources, of which many more examples have been
studied by Furuya et al. (\cite{furuya01}, \cite{furuya03}), the maser
excitation occurs on a small ($\sim 100$~AU) spatial scale and might be 
produced by the stellar jets visible in the radio-continuum. The CO-outflows
(which are also present in these sources) are either less powerful or, more 
likely, impact with a lower-density ambient molecular cloud, where conditions
are not suitable to create masers;

\item In the intermediate luminosity class, the larger energetic input from 
the CO outflow, as well as the presence of a higher-density molecular gas, 
are the main agents that determine the conditions for maser excitation 
(rather than the YSO luminosities);

\item In the most luminous sources, conditions for maser excitation are 
similar to those in the previous category, but in this case the energetic 
input is so large that all potential maser sites are excited and the 
determining factor is the YSO luminosity.

\end{enumerate}

\begin{figure}
\resizebox{\hsize}{!}{\includegraphics{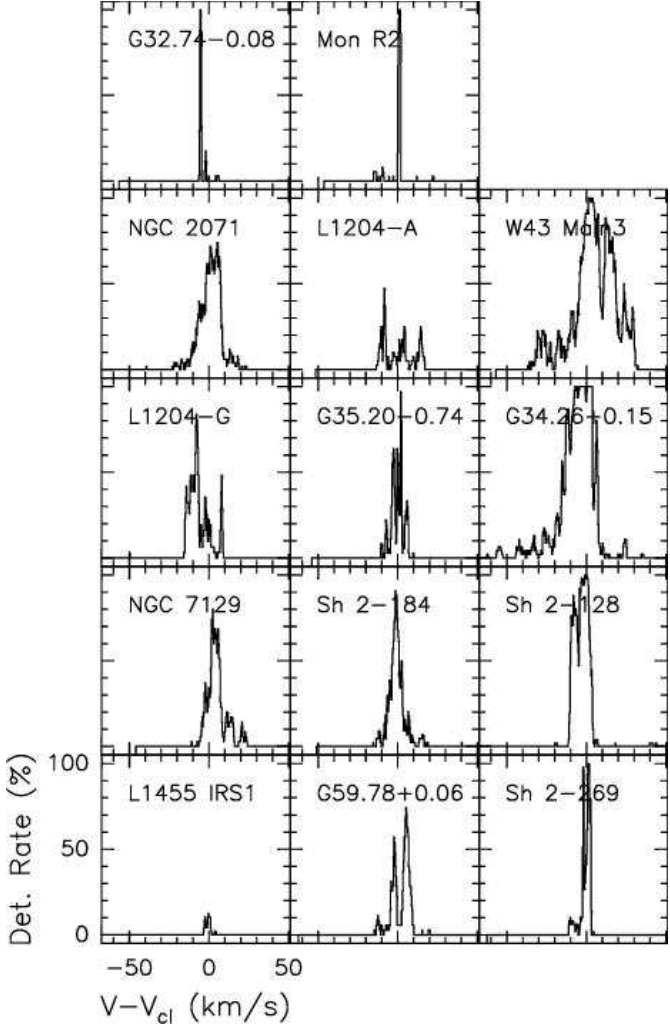}}
\caption[]{Frequency-of-occurrence histograms for all sources. All velocities
are relative to $V_{cl}$. The panels are ordered in \Lfir, which increases
from bottom to top and from left to right.}
\label{freq}
\end{figure}

\subsubsection{The scaled frequency-of-occurrence histograms}

In Fig.~\ref{freq} we show the frequency-of-occurrence histograms in order of
increasing FIR luminosity (ordered as in Fig.~\ref{grey}). The velocity 
scale is the same in all plots and the velocities are referred to 
$V_{\rm cl}$ (i.e. the quantity on the horizontal axis is $V_{\rm observed} - 
V_{\rm cl}$).

\noindent
We see that for sources with \Lfir $> 3\times 10^4$~\lsol\ the peak of the
distribution is always at 100\% (i.e. these masers are
{\it always} detected, cf. the lower envelopes discussion in the previous
subsection), and it is always at or very close to $V_{\rm cl}$.
Note that this is in agreement with the findings of Wouterloot et
al. (\cite{wou95}), who concluded that for log\Lfir $>10^{4.5}$ (i.e.
\Lfir $>3.2 \times 10^4$~\lsol) the maser detection rate is virtually 100\%.
For sources with lower values of \Lfir\ the maser is typically
detected (at the $\geq 5\sigma$ level) about 75\%-80\% of the time. Once more
the exception is L1455~IRS1, which has a detection rate at these
levels of only about 10\%. For these lower-\Lfir\ sources the peak of the
distribution of emission components can also be much further away from
$V_{\rm cl}$ than what is found for the high-\Lfir\ sources (see also 
Sect.~2.1).

\noindent
The steep decline of the histograms with velocity has already been
commented on in Sect.~2.7, as showing that the more blue- and red-shifted
maser components have shorter lifetimes than the components nearer the 
systemic (i.e. molecular cloud) velocity.

\noindent
Many of the histograms in Fig.~\ref{freq} show a main peak, and a collection
of smaller peaks (hereafter `the tail') on one side, while on the other side 
the decline is steeper.
It seems that for sources with \Lfir\ $> 3\times 10^4$~\lsol\ 
the tail is preferentially on the blue side of the main peak 
(at these \Lfir 's this is so for all 5 objects except Sh~2-128),
while at lower \Lfir\ 
the tail is either mostly on the red side 
(as in NGC7129/FIRS2, L1204-G, G32.74$-$0.08, Sh~2-184) or the histogram does 
not show these smaller peaks at all (as in the case of L1204-A, G35.20$-$0.74, 
L1455~IRS1).

\noindent
Masers and outflows are closely correlated (see e.g. Felli et
al.~\cite{FPT92}), with maser components 
at the bow shocks produced by the outflows upon impact with
the ambient medium.
Maser amplification is maximum in the plane of shocks, where the gain path
is longest (Elitzur et al.~\cite{eli89}); for high-velocity blue-shifted 
components
the plane of the shock should be perpendicular to the
line-of-sight, and therefore the gain path is rather short. However, in the
case that a well-collimated outflow is precisely aligned
with the line-of-sight, the maser can amplify the background continuum (from
the H{\sc ii}-region), and the high-velocity blue-shifted components
can become more intense. This might be the case for the tail in high-luminosity
sources.
The collimation of the outflow, the alignment of the
flow with the line-of-sight, and the H{\sc ii}-region background radio 
continuum (hence the \Lfir\ of the YSO) will determine the strength and the 
velocity offset (with respect to the molecular cloud velocity) of the 
high-velocity blue-shifted maser components. It appears that for
the sources in our sample these three effects combine most favourably in
G34.26$-$0.15 and W43~Main3 (with blue-shifted maser components up to
$\sim 60$~\kms\ and $\sim 40$~\kms\ from $V_{\rm cl}$, respectively).
According to this scenario the outflow in S~2-128 would not be aligned with
the line-of-sight, or it could have a large opening angle (or both).
The high-velocity red-shifted components (belonging to the
part of the outflow moving away from us) will always be 
weaker, because they cannot amplify the continuum background. 
If there is no outflow, or if it is driven by
a lower-luminosity YSO, high-velocity components (blue- or red-shifted) will
not be seen at all. If the maser emission comes primarily from a protostellar
disk, blue- and red-shifted components can be seen, but at velocities close
to $V_{\rm cl}$.

\noindent
The maser emission would therefore not only be a function of the luminosity
of the exciting source, but also of the geometry of the 
SFR, in particular the orientation of the  beam of the outflow.

\begin{figure}
\resizebox{\hsize}{!}{\includegraphics{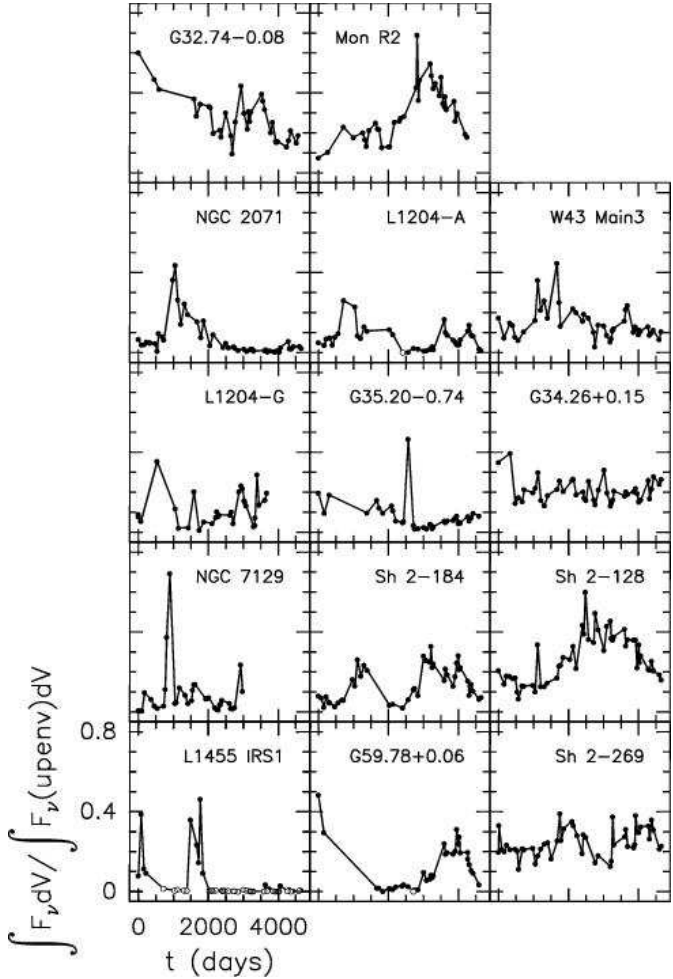}}
\caption[]{The integrated flux density, normalized by that of the upper 
envelope for each source. The panels are arranged by increasing \Lfir, which 
increases from bottom to top and from left to right. Open circles indicate 
non-detections.}
\label{fnrm}
\end{figure}

\subsubsection{The scaled integrated light-curves} 

In Fig.~\ref{fnrm} we show the light curves of the integrated flux density
$\int F {\rm d}V$ in order of increasing FIR luminosity, as in 
Fig.~\ref{grey}. 
The intensity for each source has been normalized to the 
integrated flux density of the upper envelope for that source.

\noindent
The main phenomenon that emerges from these panels is that of global long-term 
(`super') variability, which was mentioned in Sect.~2.7: in addition to the 
occasional outbursts in total integrated flux density (sometimes due to a 
single maser component in a flaring stage), several of the sources appear to 
have a more gentle, long-term variability of the total maser output. This is
most clearly seen in Sh~2-269~IRS2 and Sh~2-184, both of which show 
cyclic variations. The integrated emission of L1204-A is more fragmented and 
changes more erratically, yet also here we can see a hint of a general 
systematic variation of the total maser output
with time. For G59.78+0.06 the first 2000~days are very poorly sampled, and
it is therefore impossible to say if the broad peak we see near day 4000
has been preceded by a similar one during that earlier period. Mon~R2~IRS3,
Sh~2-128, and W43~Main3 may all show signs of global long-term variability as 
well, but with a period of the order of, or larger than, our monitoring period.
The peak in NGC2071 on the other hand, might just be an outburst with a
rapid increase in integrated flux density followed by a slow decline.
We estimate that in Sh~2-184 a full cycle is completed in about
2000~days (5.5~yrs); the same is found for Sh~2-269~IRS2, which agrees with
the $5.7 \pm 1.0$~yr derived by Lekht et al. (\cite{lekht269}). Sh~2-128 has
a cycle about twice as long (see also Lekht et al.~\cite{lekht128}).
A rough
estimate for the period in L1204-A is $\approx 3000$~days (8.2~yrs). The 
origin of this long-term cyclic variation is not known, but in the hypothesis 
that \WAT\ masers are excited behind shocks caused by the impact of 
wind-driven outflows 
with the ambient medium, it seems natural to assume that the 
variation in the total maser output is either caused by a periodic variation 
in the YSO-wind or, for sources with non-periodic variation, by the turbulent 
motions in the material of the surrounding molecular cloud.

\section{Summary}

We have used the data collected during more than 10 years of monitoring 
\WAT\ masers in 14 SFRs (Valdettaro et al.~\cite{paper1}), in order to 
extract general properties of the maser emission, and to investigate 
possible correlations between the various parameters of the maser emission 
(such as mean velocity, velocity extent, luminosity) and the FIR luminosity
of the (presumable) driving source -- usually an IRAS source. We have 
looked at velocity gradients of individual maser components, at the properties
of bursts (their duration and their increase in flux density), and at 
long-term variations in the maser output. 

\noindent
One property that comes out clearly from this analysis is the existence of a
{\it general} dependence of maser parameters on \Lfir. In addition
there are indications of the existence of different \Lfir\ regimes and a
threshold YSO-luminosity that can account for the various observed
characteristics of the \WAT\ maser emission.
We find differences in some properties of the emission of
masers associated with YSOs above and below a threshold luminosity of
around $10^4$~\lsol.
Only for sources with \Lfir $\gsim L_{\rm fir}^{\rm limit}$ a water maser is
{\it always} detected. Above $L_{\rm fir}^{\rm limit}$ one finds 
increasingly morphologically complex maser emission, from the narrow, single 
emission component in Mon~R2 to the broad, multi-component jungle of 
W43~Main3.

\noindent
The main findings of this general analysis are summarized below.

\begin{enumerate}

\item The intensity-weighted mean velocity of the maser emission
($V_{\rm up}$) is close to that of the parental the molecular cloud
($V_{\rm cl}$); their difference, normalized to the total width
($\Delta V_{\rm up}$) of the maser emission is smaller if \Lfir (YSO)
$\gsim L_{\rm fir}^{\rm limit} \gsim 7 \times 10^3$~\lsol\ (Sect.~2.1);

\item The velocity at which the maser emission is most intense, is also the
velocity where emission occurs most frequently (Sect.~2.1);

\item \Lup, the luminosity the \WAT\ maser would have if all velocity
components were to emit at their maximum level at the same time, is
correlated with the YSO FIR luminosity \Lfir\
by \Lup $= 6.37 \times 10^{-8} L_{\rm fir}^{0.81\pm 0.07}$ (Sect.~2.2);

\item The ratio between \Lmax, the maximum \WAT\ luminosity measured during 
the entire monitoring period, and \Lup\ ranges between 0.25 (Sh~2-269~IRS2) 
and 0.80 (Mon~R2~IRS3), but does not depend on \Lfir\ (Sect.~2.2);

\item The anti-correlation between \Lfir\ and the ratio of the maximum and the 
mean integrated flux density indicates that high-luminosity sources tend to be 
associated with more stable masers, while lower luminosity ones have more 
variable emission (Sect.~2.3);

\item Higher maser power goes into more emission channels that are spread over
a larger range in velocity. While in a diagram of
$\Delta V_{\rm up}$ versus \Lfir\ only an upper envelope can be defined,
there is a real correlation between both the number of components and the
maximum velocity extent of the maser emission and the luminosity of the
\WAT\ maser if all are derived from the same spectrum (Sects.~2.4, 2.5);

\item For $\sim 15$ relatively isolated maser components we have made an 
analysis
(using Gaussian-fits) of the emission velocity. We find both acceleration
and deceleration in equal numbers, with gradients that range between
0.02 and 1.8~\kmsyr. The smaller value refers to a component that could be
traced over the whole 4600~days of monitoring; the higher value was found
for a burst-component of $\sim 60$~day duration (Sect.~2.6);

\item We have looked in some detail at 14 outbursts in 9 emission components
in 6 sources. The shortest/longest duration is $\sim 60/900$~days. Bursts of
shorter duration are not detectable due to our sampling interval
(typically~2-3 months), while those of longer duration are usually not
recognized as a burst. The increase in flux density during a burst ranged
from 40\% to $\gsim 1840$\%; the largest absolute increase was found in
Mon~R2 ($\sim 820$~Jy) (Sect.~2.7);

\item In several sources we find a complete (Sh~2-269~IRS2 and Sh~2-184) or
partial (Mon~R2~IRS3, Sh~2-128, and W43~Main3) cycle of integrated flux
density changes over long timescales.
For the first two sources we find this period to be of the order of 5.5~yrs.
Also L1204-A may show such a long-term variation; in this case a rough
estimate indicates a period of $\sim 8$~yrs (Sect.~2.8.5);

\item From the velocity-time-intensity plots we identify a limiting \Lfir\ 
of $\sim 3 \times 10^4$~\lsol. For sources with \Lfir\ above this limit
the maser emission becomes increasingly structured and more extended in 
velocity with increasing \Lfir. Below this limit the maser emission shows
the same variety of morphologies, but without a clear dependence on \Lfir\ 
(Sect.~2.8.1);

\item From the lower envelopes we can identify the same limiting value of
\Lfir $\approx 3\times 10^4$~\lsol: all sources above this limit have 
at least one maser component that is {\it always} present, at a level of 
$1-20$\% of 
the peak flux density in the upper envelope spectrum, and with a velocity 
very close to that of the upper envelope peak and to the 
molecular cloud velocity (Sect.~2.8.3);

\item From the upper envelopes we deduce the possible existence of three
regimes of maser excitation, associated with three ranges in YSO \Lfir.
For the lowest ranges, \Lfir $<  4 \times 10^2$~\lsol\ and
$4 \times 10^2$~\lsol $\lsim$ \Lfir $\lsim  6 \times 10^4$~\lsol, the maser
excitation depends mostly on the strength of the outflow and the density of
the surrounding molecular cloud, while for \Lfir $\gsim  6 \times 10^4$~\lsol\
the YSO-luminosity is the determining factor (Sect.~2.8.3);

\item The frequency-of-occurrence histograms show that
\Lfir\ $\approx  3\times 10^4$~\lsol\ is a threshold value for the FIR
luminosity of the presumed maser driving-source. For sources with \Lfir\
above this limit the peak of the distribution is always at 100\% (i.e. the
maser is always detected). Below it, the typical detection rate (at the
$>5 \sigma$-level) $75-80$\%. The exception is L1455~IRS1, which has a
detection rate of only $\sim 10$\% (Sect.~2.8.3). 
There is also a lower bound to \Lfir\ (at least $\lsim 430$~\lsol),
below which the associated maser source is not detectable most of the time 
(Sect.~2.8.1).

\item The presence or absence of blue-shifted high-velocity maser components
in the frequency-of-occurrence histograms led us to conclude that the 
maser emission is a function of not only the luminosity of the YSO, but 
also of 
the beaming properties of the outflow with respect to the observer 
(Sect.~2.8.4).

\end{enumerate}

\acknowledgements
This paper was written in spite of the continued efforts by the Italian
government to dismantle publicly-funded fundamental research in general and 
the C.N.R. in particular.\hfill\break\noindent
The 32-m VLBI antenna at Medicina is operated by the Istituto di 
Radioastronomia of the Consiglio Nazionale delle Ricerche, in Bologna.

\noindent
This research has made use of NASA's Astrophysics Data System Bibliographic 
Services

\end{document}